\documentclass[twocolumn]{aa}
\usepackage{graphicx}
\usepackage{url}
\usepackage{natbib}
\usepackage{txfonts}
\usepackage[normalem]{ulem}
\usepackage{longtable}
\usepackage{color}
\usepackage{amssymb,xcolor}
\usepackage{amsmath}
\newcommand{\swift}  {Swift~J0243.6+6124}

\def\msun{~{\rm M}_\odot}
\def\rsun{~{\rm R}_\odot}
\def\ergs{erg/s}
\def\esca{$erg\, s^{-1}\, cm^{-2}\, \AA^{-1}$}

\begin{document}

   \title{Unveiling the origin of the optical/UV emission from the Galactic ULX Swift\,J0243.6+6124 during its 2017--2018 giant outburst}

   \subtitle{}
  \author{J. Alfonso-Garzón \inst{1} \and J. van den Eijnden \inst{2} \and N. P. M. Kuin \inst{3}  \and F. F\"{u}rst \inst{4} \and A. Rouco-Escorial \inst{5} \and J. Fabregat \inst{6} \and \\ P. Reig \inst{7,8} \and J. M. Mas-Hesse \inst{1} \and P. A. Jenke \inst{9} \and C. Malacaria \inst{10} \and C. Wilson-Hodge \inst{11}  
}

\authorrunning{}
\titlerunning{\swift}

   \offprints{}

    \institute{Centro de Astrobiolog\'{\i}a (CAB), INTA--CSIC, ESAC, Camino Bajo del Castillo s/n, 28692 Villanueva de la Ca\~nada, Spain \\ \email{julia@cab.inta-csic.es}
	\and Department of Physics, University of Warwick, Coventry CV4 7AL, UK
	\and Mullard Space Science Laboratory, Holmbury St Mary, Dorking, Surrey RH5 6NT, UK
	\and Quasar Science Resources SL for ESA, European Space Astronomy Centre (ESAC), 28692 Villanueva de la Can\~ada, Madrid, Spain 
	\and European Space Agency (ESA), European Space Astronomy Centre (ESAC), Camino Bajo del Castillo s/n, 28692 Villanueva de la Ca\~nada, Spain
	\and Observatorio Astron\'omico, Universidad de Valencia, Catedr\'atico Jos\'e Beltr\'an 2, 46980 Paterna, Spain 
	\and Institute of Astrophysics, Foundation for Research and Technology-Hellas, 71110 Heraklion, Greece 
	\and Physics Department, University of Crete, 71003 Heraklion, Greece
	\and University of Alabama in Huntsville, Huntsville, AL 35805, USA  \and International Space Science Institute, Hallerstrasse 6, 3012 Bern, Switzerland 
        \and ST12 Astrophysics Branch, NASA Marshall Space Flight Center, Huntsville, AL 35812, USA
        }
   \date{Received; Accepted}

\abstract
{From late September 2017 to February 2018, the Be X-ray binary (BeXB) Swift~J0243.6+6124 underwent an unprecedently bright giant outburst. The reported X-ray luminosities were so high that the system was classified as an Ultraluminous X-ray source (ULX). It was also the first BeXB pulsar showing radio jet emission. The source was not only bright in X-rays and radio, but also in optical and UV wavelenghts. } 
{In this paper we aim to understand the origin of the observed optical/UV fluxes simultaneous to the X-ray emission.} 
{We studied the optical/UV light curves in comparison with the X-ray fluxes along the outburst. We considered the main mechanisms that can explain the optical/UV emission in X-ray binaries. Due to the tight correlation observed between the optical/UV and X-ray light curves, reprocessing of X-rays seems to be the most plausible explanation. We calculated the timescales of the light curves decays and studied the correlation indexes between the optical and X-ray emission. Finally, we built a physical model considering X-ray heating of the surface of the donor star, irradiation of the accretion disk, and emission from a viscously heated accretion disk, in order to reproduce the observed optical/UV SEDs along the outburst. We considered in our model that the Be circumstellar disk was co-planar to the orbit, and then we neglected its irradiation in the current model. As an input of the model, we used as incident X-ray luminosities those calculated from the bolometric X-ray fluxes obtained from the spectral fit of the \textit{Swift}/XRT and BAT observations.}
{The timescales of the exponential decay of the outburst are 2--4 times longer for the UV and optical light curves than for the X-ray light curve. The correlation index between the optical/UV and X-ray fluxes varies between optical/UV filters and when different X-ray bands are considered, and is larger for the rise than for the decay phase of the outburst for the fluxes at redder wavelengths. The modelling of the SED shows that X-ray heating of the companion star surface is the main mechanism contributing to the UV emission and contributes significantly to the optical emission during the whole outburst. X-ray irradiation of the accretion disk is necessary to reproduce the optical observed fluxes from MJD~58047 to 58120, and contributes significantly to the UV fluxes close to the peak of the outburst. As a first attempt, the fits yield an increasing value of the outer radius of the accretion disk along the outburst. An alternative interpretation points to variations on the geometry of the inner flow and the fraction of reprocessed X-ray emission during the outburst. On the other hand, variations on the geometry of the Be circumstellar disk could also play a role, but they have not been considered in the current model.}
{Reprocessing of X-rays as X-ray heating of the Be star surface and irradiation of the accretion disk are the two main mechanisms that can reproduce the observed optical/UV emission during the 2017--2018 giant outburst of \swift.
}


   \maketitle

\section{Introduction}
\label{sec:int}

Be/X-ray binaries (BeXBs) are systems formed by a neutron star (NS) and a Be star. The Be star expels matter from the photosphere to create a quasi-Keplerian disk around its equatorial plane, and emission lines and strong infrared excesses are usually observed in these systems due to the presence of this circumstellar disk \citep{Reig2011}. BeXBs can exhibit two types of X-ray outbursts: type I or normal outbursts, with moderate intensities (L$_{X}\approx$ 10$^{36}$--10$^{37}$\,erg\,s$^{-1}$); and type II or giant outbursts, with L$_{X} >$~10$^{37}$\,erg\,s$^{-1}$. Type I or normal outbursts take place close to the periastron passage of the NS, when the NS can cross the circumstellar disk and accrete material from it emitting in X-rays. Type II outbursts can occur at any orbital phase, with a duration usually longer than the orbital period, and their origin remains unclear, although two main scenarios have been proposed to explain their ocurrence. In one hand, type II outbursts could take place due to the interaction of the NS with a tidally warped and precessing circumstellar disk \citep{moritani11,moritani13,okazaki13}. On the other hand, for highly misaligned circumstellar disks, giant outbursts could take place when the Be disk undergoes strong eccentricity growth due to Kozai-Lidov oscillations and the material overflows the Roche Lobe and is captured by the NS \citep{martin14a,laplace2017,Martin2019}.

Swift~J0243.6+6124 is a unique system, since it is the first and only Ultraluminous X-ray Pulsar (ULXP) in our Galaxy \citep{doroshenko18}. It was discovered by the Burst Alert Telescope (BAT) on board the Neil Gehrels {\it Swift} observatory on 3 October 2017 at the beginning of a giant X-ray outburst \citep{kennea17}. A spin period of around 9.86\,s was detected by \textit{Swift}/XRT \citep{kennea17}, Fermi/GBM \citep{jenke17}, and NuSTAR \citep{bahramian17}. The Fermi/GBM observations allowed accurate determination of the orbital period P$_{orb}$=27.587\,days and eccentricity $e$=0.098 \citep{Wilson-Hodge2018, zhang19}. 

A few days after the discovery of the system, \citet{kouroubatzakis17} obtained an optical spectrum of the source and claimed it was a new Be/X-ray binary (BeXB). A detailed analysis of the spectroscopic and photometric observations of the optical counterpart of \swift\ was performed by \citet{Reig2020} (hereafter Paper~I), who classified the optical companion of this system as an O9.5Ve star. In this work, long-term optical variability was observed in the available photometric observations, which was interpreted as variations in the Be circumstellar disk.

Different estimations of the distance to the source have been obtained. \citet{bikmaev17} proposed a distance of $\sim 2$ kpc based on optical photometric observations. \citet{doroshenko18} and \citet{zhang19} proposed a lower limit of $\sim 5$\,kpc based on the spin-up rate and accretion-torque models. An estimation of 4.8$\pm$0.5\,kpc was obtained from optical photometric observations in Paper~I. It should be stressed that many papers focused on the X-ray analysis of the 2017 giant outburst of \swift\, have considered in their analyses the distance estimated from the Gaia Data Release 2 at face value, which is d=6.8\,kpc \citep{bailer-jones18}. However, recent estimations from the Gaia Early Data Release 3 points to a distance value of d=5.2$\pm$0.3\,kpc \citep{Bailer-Jones2021}, in agreement with the estimation from the optical data in Paper~I. We will use this value in our calculations in this work. This value of the distance implies X-ray luminosities at the peak of the outburst close to L$_{X}\sim$ 10$^{39}$ \ergs, which still indicates that the Eddington limit for the neutron star was exceeded during the outburst, and the system can still be classified as an Ultraluminous X-ray Source (ULX).

Several studies have been focused on the analysis of the X-ray spectrum, which has been characterized by a cut-off power-law continuum \citep{zhang19} with an evolving iron line emission \citep{jaisawal19}. Spectral and pulse profile analyses have been performed by \citet{Wilson-Hodge2018, Doroshenko2020} and \citet{Liu2022}. There have been also many studies focused on studying the configuration of the magnetic field of \swift\,\citep{doroshenko18, Wilson-Hodge2018,tsygankov18, Doroshenko2020, Bykov2022, Serim2023}, and a cyclotron line from 120 to 146\,keV was proposed by \citet{Kong2022}. One or more blackbody components associated with the hot spot around the polar region and thermal emission from the accretion column and from the photosphere of a possible ultrafast outflow have been proposed to fit the X-ray spectrum at lower energies \citep{tao19,vandenEijnden2019b}.

An interesting and surprising feature of the source was the detection of radio emission during the X-ray outburst, which has been interpreted as coming from a jet \citep{vandenEijnden2018}. Radio emission was not expected for any BeXB, since it was believed that strong magnetic fields should inhibit the jet formation \citep{Fender2000, Massi2008}. Anyway, the radio emission detected in \swift\ was two orders of magnitude fainter than for other X-ray pulsars in low-mass X-ray binaries (LMXBs) at similar X-ray luminosities, and this difference could be due to the strong magnetic field. During a later X-ray re-brightening of the source at significantly lower X-ray luminosity, the radio emission re-brightened to similar levels as the main outburst peak \citep{vandenEijnden2019a}.

In Paper~I, we identified a very bright optical outburst in the $V$--Johnson band optical light curve of \swift, taking place simultaneously with the X-ray emission. In this paper, we explore the possible mechanisms that can explain the optical and UV emission observed during the giant outburst of the system and the correlation with the X-ray emission. Some of these mechanisms involve variations in the companion star and/or the circumstellar disk of the Be star and can produce optical variability of tenths of magnitudes. They comprise different scenarios such as ejections of matter from the Be star \citep{hubert98, mennickent02}, precession and/or warping or changes in the inclination of the circumstellar disk \citep{okazaki13, martin14a, Martin2019}, and formation and dissipation of the circumstellar disk (see for example \citealt{Reig2023}).

On the other hand, there are other mechanisms directly related to the X-ray emission. The correlation indexes between the X-ray and optical/IR fluxes have been traditionally used in the past to study the origin of the optical emission in X-ray binaries. \citet{vanParadijs1994} affirmed that if the optical emission comes from reprocessing of X-rays in the accretion disk of LMXBs, then $L_{\mathrm{V}}\propto T^{2} \propto \,L_{\mathrm{X}}^{0.5} a$, where $T$ is the effective temperature of the irradiated accretion disk and $a$ is the orbital separation. Statistical studies have identified the emission from irradiated disks as the main mechanism contributing to the optical and IR light in many LMXBs \citep{Russell2006}. For neutron star LMXBs, the origin of the optical/IR emission has been found to vary with X-ray luminosity and source type (atolls, millisecond X-ray pulsars, and Z sources). \citet{russell07} concluded that X-ray reprocessing can explain the optical/IR emission in Z sources at all luminosities. For atoll and millisecond pulsars this effect would dominate at low luminosities ($L_X\sim 10^{36}$ \ergs\ and $L_X\sim 10^{37}$ \ergs\,respectively), while above $L_X\gtrapprox  10^{37}$ \ergs, optically thin synchrotron emission from the jet would dominate the NIR light \citep{russell07}.

\citet{King1998} predicted that when the optical emission comes mainly from an irradiated disk, the exponential decay timescales of the optical and IR light curves should be roughly $\sim$2--4 times the e-folding time of the X-ray light curve. This relationship has been observed in several X-ray novae, with typical X-ray light curve decay timescales of 30 days \citep{Chen1997}, and more recently also for the black hole candidate X-ray transient XTE J181--330 \citep{Rykoff2007}.

Statistical studies are useful to study the general behaviour of X-ray binaries and to find patterns for different types and subtypes, or when the sources are at different spectral states or luminosities. However, simple flux-flux correlations can be too simplistic to understand in detail the effects of reprocessing of X-rays into longer wavelengths for an individual system, and more sophisticated modelling is needed to explain the origin of the optical/UV emission during outbursts \citep{Vrtilek1990, Gierlinski2009}. 

On the other hand, these studies have been traditionally performed for LMXBs, since for high-mass X-ray binaries (HMXBs) the emission from the optical companion is expected to dominate over reprocessing effects at optical and ultra-violet (UV) wavelengths. Indeed, such bright optical outbursts taking place simultaneously with the X-ray outbursts are not common in BeXBs. As far as we know, the only system displaying very bright optical outbursts during active X-ray epochs that has been reported, is the Be-ray binary 1A\,0538--66 (this source also reached L$_{X}\sim$ 10$^{39}$ \ergs), which can reach amplitudes of the variability of $\sim$2\,mag in the $V$--band \citep{charles1983}. Several mechanisms have been proposed to explain the extremely bright optical outbursts of this system, including irradiation inside the binary system, although this system showed many other peculiarities with respect to other BeXBs, which complicated the interpretation of the observations \citep{charles1983}. \citet{ducci19} performed a very detailed analysis of the optical emission of this system, considering both irradiation of the accretion disk and heating of the surface of the Be star, but concluded that in the case of 1A\,0538--66, the optical emission is mainly powered by reprocessing of X-rays in the envelope surrounding the binary system. 

In order to constrain the origin of the optical/UV emission, there are several methods that can be used:
\begin{itemize}
    \item Compare the light curves in the different optical/UV bands with the X-ray light curve.
    \item Study the correlation between the measured optical/UV and the X-ray fluxes.
    \item Measure the timescales of the exponential decay of the outburst for the light curves in each band.
    \item Develope physical models in order to reproduce the observations.
\end{itemize}

In this paper, we have carried out a detailed analysis of the optical and UV emission from \swift, in correlation with the X-ray fluxes during the 2017--2018 giant outburst, performing the calculations mentioned above. In Sect.\,\ref{sec:obs}, we describe the observations used on this analysis. In Sect.\,\ref{sec:anal}, we present the calculations and data analysis. In Sect.\,\ref{sec:disc}, we discuss the possible origins of this emission and describe the physical model we propose to reproduce the optical/UV emission. Finally, in Sect.\,\ref{sec:sum}, we summarize the main results and present the conclusions of the paper.

\section{Observations}
\label{sec:obs}
\subsection{Optical/UV photometry} \label{sec:optical_photometry}

The optical $V$--Johnson band light curve contains data from several observatories (see \citealt{Reig2020} for a detailed description of the optical data). We included the light curve from the ASAS--SN Variable Stars Database \citep{shappee14,jayasinghe19}, correcting the data from the contamination due to the presence of a close star in the photometric extraction aperture. Photometric observations from Skinakas and Aras de los Olmos observatories are also included. There are also some observations from the American Association of Variable Star Astronomers (AAVSO; \citealt{Watson2012}). Additionally, we have also included the photometric observations in the $v$ filter from the Ultra-Violet/Optical Telescope (UVOT, \citealt{Roming2005}) on board the \textit{Neil Gehrels Swift Observatory} \citep[hereafter \textit{Swift};][]{Gehrels2004} instrument. To convert the UVOT-$v$ filter measurements to the Johnson $V$--band we have applied a filter conversion correction. This correction has been calculated by comparing the synthetic photometry of the theoretical spectrum of an O9.5V star reddened with E(B-V)=1.20\,mag \citep{Reig2020} in both filters.  

\begin{figure*}
\begin{center}
\resizebox{\hsize}{!}{\includegraphics{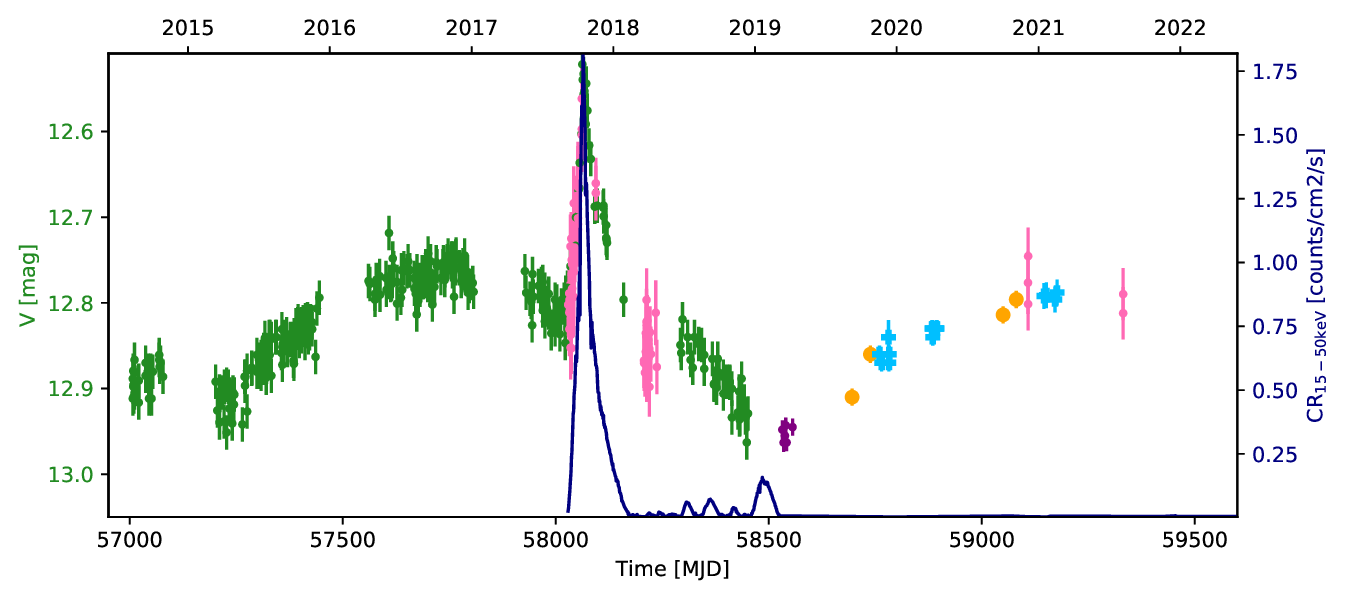}}
\caption[]{Optical vs X-ray long-term light curves of \swift.  the light curve contains observations from different sites: green points come from ASAS--SN Variable Stars Database, pink points correspond to converted \textit{Swift}/UVOT--$v$ observations, purple points are from AAVSO, orange points are from Skinakas observatory, and light blue points are from Aras de los Olmos Observatory (see Sect.\,\ref{sec:optical_photometry} for details). The navy blue line represents the X-ray count rate light curve measured with \textit{Swift}/BAT in the 15-50\,keV band. }
\label{fig:VLxLC_lt}
\end{center}
\end{figure*}

The long-term optical light curve of \swift\, in the $V$--Johnson band, together with the \textit{Swift}/BAT (15--50\,keV) light curve, is shown in Fig.~\ref{fig:VLxLC_lt}.

Multi-colour photometric observations from the other UVOT filters have also been analysed in this work. The UV and optical light curves of \swift\, were obtained after checking the images using the uvotsource ftool to do aperture photometry. Data where the source fell on a low-sensitivity patch have not been included.

\subsection{\textit{Swift}/XRT and BAT observations}

The \textit{Swift} observatory monitored the outburst of Swift J0243.6+6124 in detail. In order to characterise the X-ray evolution throughout the outburst, we used the data from both the X-ray Telescope \citep[XRT;][]{burrow2005} and the Burst Alert Telescope \citep[BAT;][]{barthelmy2005}. The XRT performed pointed observations of the outburst at a variable but high (up to daily) cadence, changing between photon counting (PC) and window timing (WT) modes throughout the outburst and its re-brightening events dependent on the target's brightness. A large fraction of the observations was taken in automatic mode, accumulating the majority of data in the suitable mode for the source flux levels. We extracted XRT X-ray target and background spectra, as well as response and ancillary files, using the \textit{Swift} online data reduction pipeline \citep[][\url{www.swift.ac.uk/user_objects/}]{evans2009} for all observations of the 2017/2018 outburst. This pipeline automatically accounts for the pile-up during the brightest phases of the outburst. For \textit{Swift}/BAT, we downloaded the daily source count rates from the Hard X-ray Transient Monitor webpage \citep[][\url{https://swift.gsfc.nasa.gov/results/transients/}]{krimm2013}. We note that, compared to the BAT light curves as reported during the outburst, any anomalously low count rates due to saturation during the outburst peak have been corrected in the current version of the monitor webpage.

\section{Data analysis}
\label{sec:anal}
In this section, we present the methods that we have used to convert from observed photometry to monochromatic fluxes, in order to build the Spectral Energy Distributions (SEDs) corresponding to each time. We also present a new estimation of the interstellar reddening. After that, we describe the method to isolate the extra emission intrinsic to the outburst. The method used to estimate the bolometric X-ray fluxes is also described.
Moreover, in order to compare with previous multi-wavelength analysis (see Sect\,\ref{sec:int}), we calculate the timescale of the exponential decay in the different bands, and the F$_{X}$/F$_{opt/UV}$ correlation indexes.

\subsection{Transformation from observed magnitudes to monochromatic fluxes}
\begin{figure}
\begin{center}
\resizebox{\hsize}{!}{\includegraphics{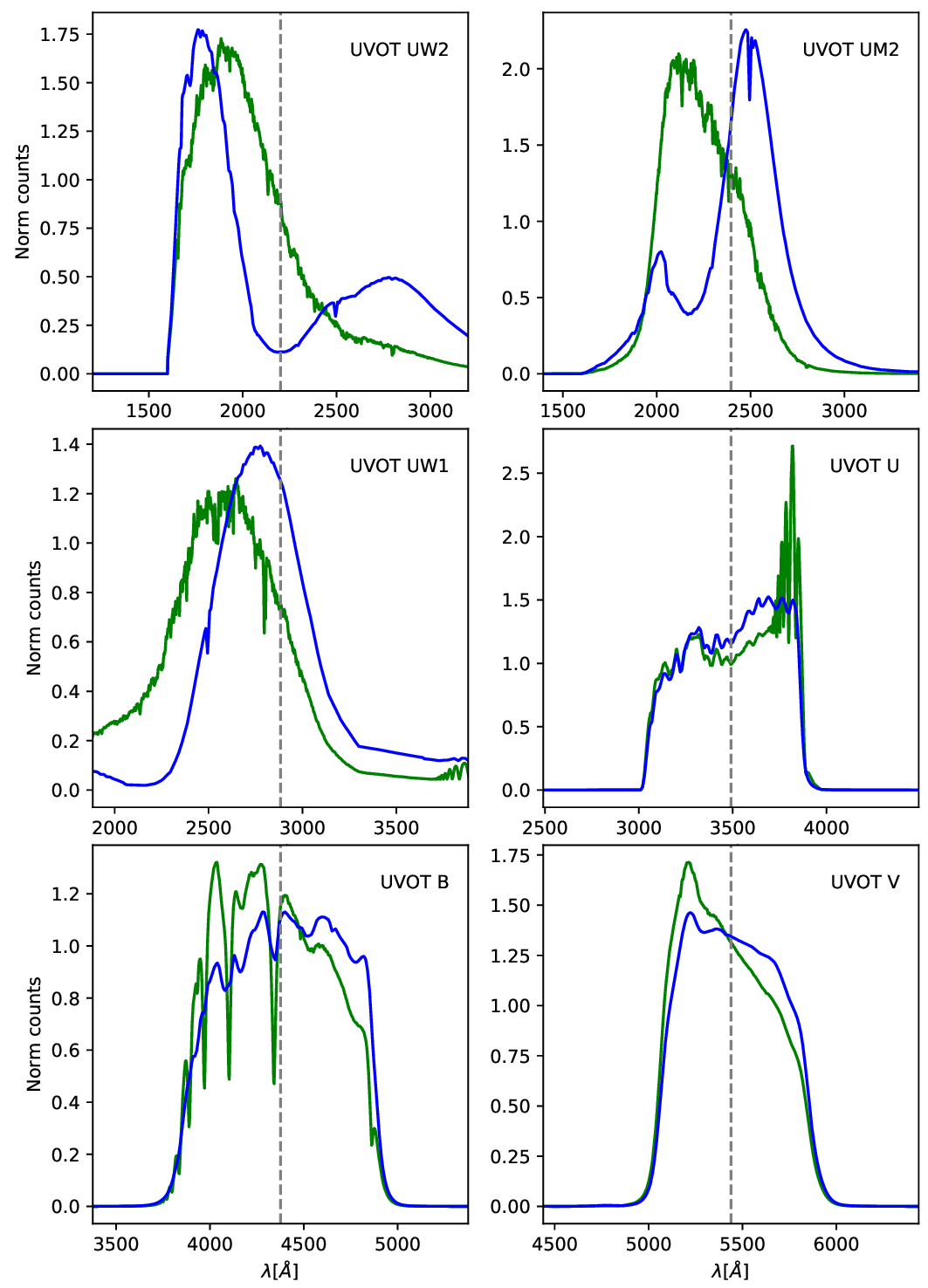}}
\caption[]{The normalized transmitted counts through the \textit{Swift}/UVOT filters for Vega (green) and for the theoretical model of an O9.5~V star reddened with E(B-V)=1.20\,mag (blue). They have been obtained multiplying the flux density by the effective area of each filter and converting fluxes to count rates. The effective wavelengths for each filter are marked with gray dashed lines. The largest differences are found for the UVOT-$uw1$ and UVOT-$uw2$ filters.}
\label{fig:FCs}
\end{center}
\end{figure}

To convert the observations in the $V$--Johnson band to monochromatic fluxes, we used the standard zero point method. The UVOT magnitudes are provided in the AB-photometric system. In order to convert them to monochromatic fluxes and be able to use these measurements to build the SEDs, we calculated the corrections to the filter response depending on the spectral shape of the observed object, as described in \citet{Brown2016}. As these authors explain, the monochromatic fluxes provided by the pipeline are not always correct, since these values are obtained using the zero points corresponding to Vega and this difference can be crucial at the UV wavelengths. We calculated the flux correction (FC) for each filter, considering the theoretical spectrum of an O9.5V star, and reddening of E(B-V)=1.20\,mag (see Sect.\,\ref{sub:redd}). The visualization of the differences between considering Vega or the theoretical reddened spectrum of \swift\, can be observed in Fig.\ref{fig:FCs}. The largest differences are found for the UVOT-$uw1$ and UVOT-$uw2$ filters.

\subsection{Estimation of the interstellar reddening}
\label{sub:redd}
In order to identify the contribution of each emission mechanism to the total optical and UV observed light, correctly determining the reddening is needed. 

In Paper~I, we obtained the spectral type of the optical companion to be an O9.5~V star, and we provided two independent measurements of the extinction: from optical photometric observations we obtained E(B-V)=1.24$\pm$0.02\,mag, while when we considered the strength of the diffuse interstellar bands (DIBs) we got E(B-V)=1.1$\pm$0.2\,mag. As we mentioned in Paper I, the determination of the interstellar extinction in Be stars (and in BeXBs) can be affected by the presence of the circumstellar disk around the Be star, which leads to an excess in the IR and redder optical wavelengths. We present here a new determination of the interstellar extinction, obtained by fitting the observed Spectral Energy Distribution (SED) points, including also the UV data from \textit{Swift}/UVOT, with synthetic photometry calculated from a grid of reddened theoretical spectra. 

\begin{figure}
\begin{center}
\resizebox{\hsize}{!}{\includegraphics{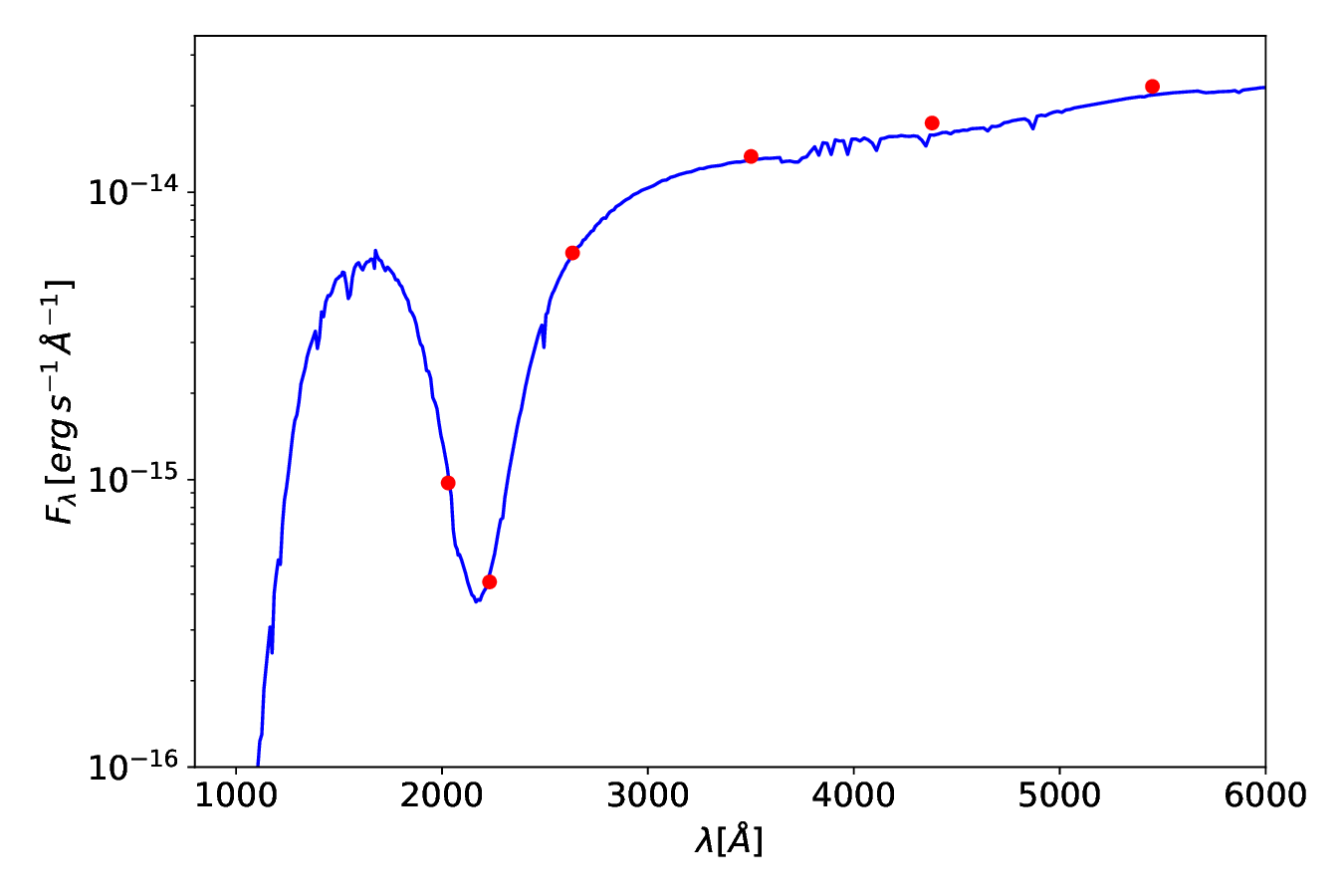}}
\caption[]{The observed fluxes of \swift\,in quiescence are marked with red stars. The reddened theoretical spectrum corresponding to an O9.5~V star, considering a distance of d=5.2\,kpc, is plotted in blue. The best fit is a value of E(B-V)=1.20$\pm$0.01\,mag.}
\label{fig:reddening}
\end{center}
\end{figure}

In order to minimize the contribution from the circumstellar disk and the emission related to the X-ray outburst, we selected the faintest observations from the long-term light curves, obtained in the least active X-ray epoch (close to the epoch around MJD 58530, see Fig.\,\ref{fig:VLxLC_lt}). In this case, we would only expect to observe the contribution of the Be star and the circumstellar disk. From the long-term variability of the optical light curves, and from the values of the equivalent width (EW) of the H$\alpha$ line, we expect that the contribution from the Be circumstellar disk is minimum at that epoch \citep{Reig2020, Liu2022b}.
For the SED fit, we built a grid of values of E(B-V) from 0.9 to 1.4 with step 0.01, and used the theoretical spectra of a O9.5~V star of T$_{eff}$=32000~k, log g = 4.0 and Z=Z$\sun$ from the Castelli \& Kurucz 2004 Stellar Atmosphere Models \citep{Castelli2003}, assuming the extinction law from \citet{Fitzpatrick1999}. We considered a distance of 5.2\,kpc and obtained synthetic photometry to compare with the observations. We got the best fit for E(B-V)=1.20$\pm$0.01\,mag, consistent with the determination in Paper~I. Similar results are obtained if the SED fit is performed only to the observed fluxes in the U+UV filters (excluding the B, and V filters, which would be more affected by the contribution of the circumstellar disk). The results of the SED fit are shown in Fig.\,\ref{fig:reddening}.

\begin{figure}
\begin{center}
\resizebox{\hsize}{!}{\includegraphics[width=1.1\textwidth]{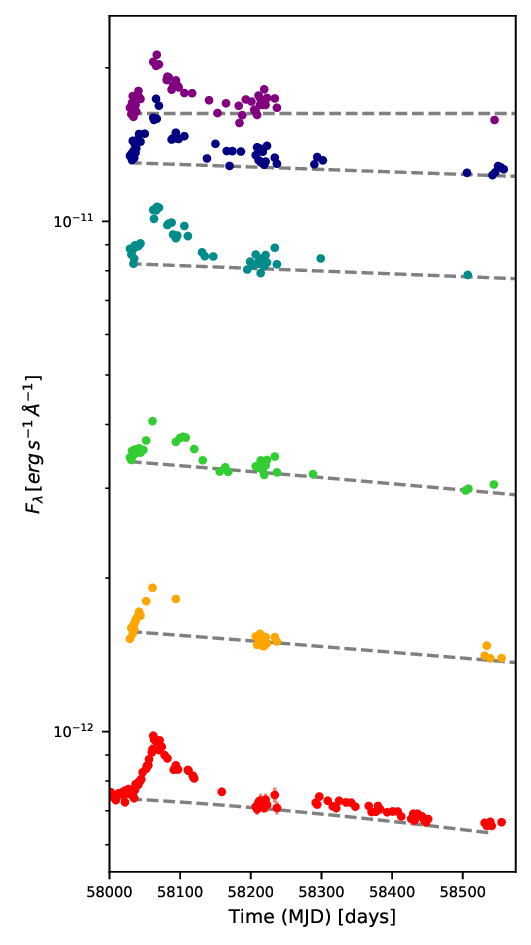}}
\caption[]{Light curves of the de-reddened monochromatic fluxes in the optical and UV light curves of \swift. From top to botton: UVOT-$uw2$, UVOT-$um2$, UVOT-$uw1$, UVOT-$u$, UVOT-$b$, and Johnson $V$. Monochromatic fluxes have been dereddened with E(B-V)=1.20\,mag. The dashed gray lines represent the fits used to model the Be+circumstellar disk contribution (see Sect.\,\ref{sub:detrend}).}
\label{fig:optUVLCs}
\end{center}
\end{figure}

The de-reddened multi-colour UVOT light curves together with the combined observations in the $V$-Johnson band, all in monochromatic fluxes and corrected from the estimated reddening, are shown in Fig.~\ref{fig:optUVLCs}.

\subsection{Isolation of the extra optical/UV emission due to the outburst}
\label{sub:detrend}

As we mentioned above, clear long-term variability can be observed in the $V$-Johnson band light curve shown in Fig.~\ref{fig:VLxLC_lt}. This oscillation has a quasi-period of $\sim$1200\,days. Similar variability over a period of $\sim$1000\,d was reported in the past by \citealt{Nesci2017} from the analysis of the B UCAC4 light curve, and by \citealt{Stanek2017} from the ASAS-Sn Sky Patrol light curve. Same type of variability was observed in the WISE light curves (Paper~I), being the amplitude of the variability larger for longer wavelengths.

\begin{figure}[h]
\begin{center}
\resizebox{\hsize}{!}{\includegraphics{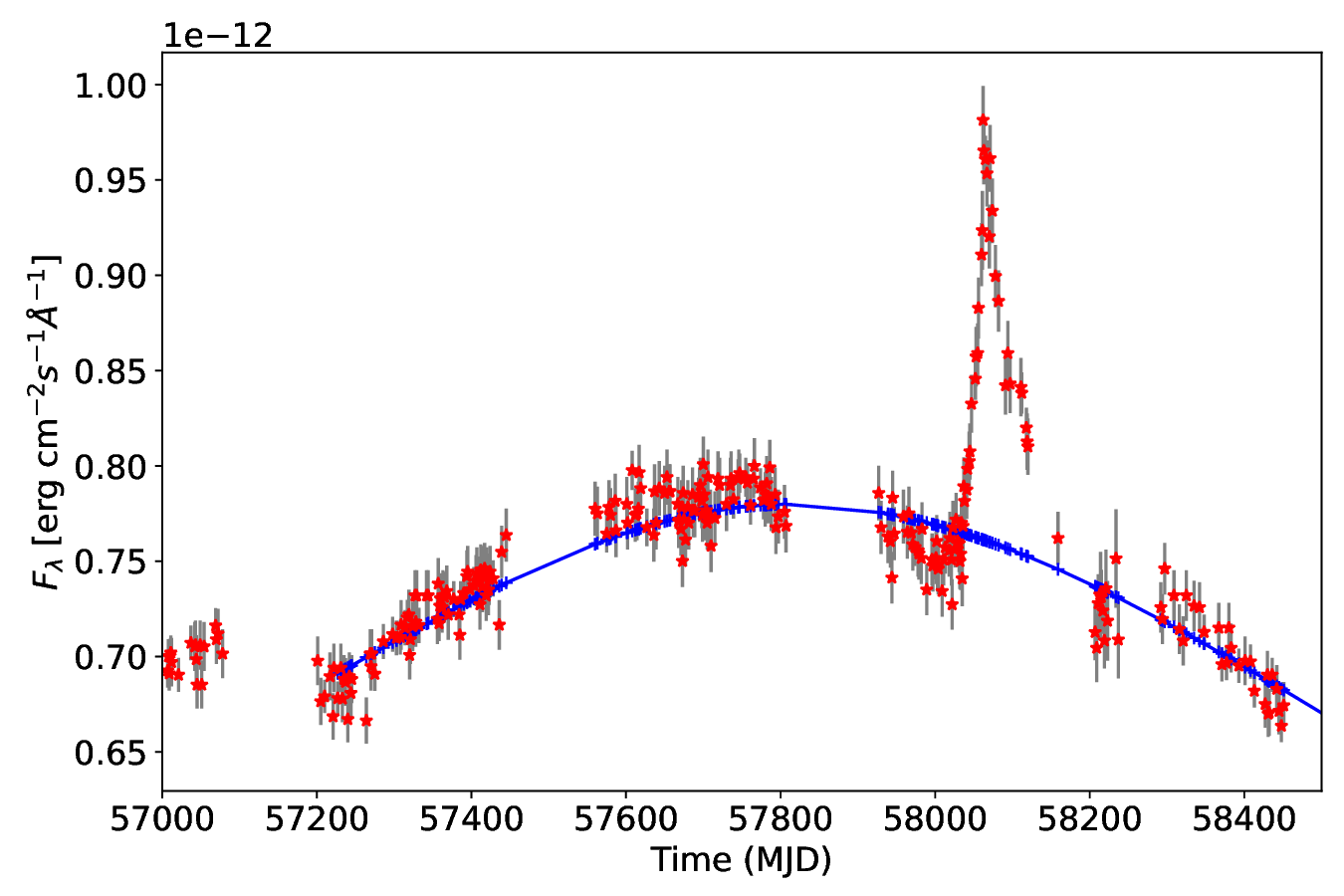}}
\caption[]{Fit of the long-term variations of the observed fluxes in the optical $V$--Johnson band. These estimated fluxes (in blue) have been removed from the observed fluxes (in red) to estimate the extra emission due to the outburst (see Sect.\,\ref{sub:detrend}).}
\label{fig:ltVfit}
\end{center}
\end{figure}

These modulations can be explained by the replenishment and dissipation of the circumstellar disk, and this explanation is in agreement with the evolution of the EW(H$\alpha$) line \citep{Reig2020, Liu2022b}. In order to have an estimation of the emission from the Be star and the circumstellar disk in the $V$--Johnson band, we performed a cubic spline fit to the points outside the outburst (we did not consider the points between MJD 57850 and 58300 MJD) on the long-term $V$--Johnson light curve (see Fig.\,\ref{fig:ltVfit}). 
For the other optical and UV bands for which the temporal coverage was poorer, we just performed a linear fit to those points out of the outburst between MJD~58200--58300 (epoch when it is considered that the giant outburst ended) and MJD 58500--58600 (epoch when the minimum fluxes were measured at all the wavelengths). Then we substracted this linear fit to the observed fluxes along the outburst. 
The long-term light curves of the fluxes in the six UV/optical filters are shown in Fig.\ref{fig:optUVLCs}. The estimation of the Be+disk emission at each wavelength is plotted with a gray dashed line. The variations due to changes in the circumstellar disk are larger for redder wavelengths (as can be seen from the slopes of the fits in Fig.\,\ref{fig:optUVLCs}).

We used these residuals (after substracting the Be+disk component) to build the daily optical/UV SEDs that trace the emission exclusively due to the outburst. These SEDs will be shown and studied in Sect.\,\ref{sub:SEDfits}.

\subsection{Estimation of the X-ray fluxes}
\label{sub:estX}
We fitted each XRT spectrum with a combined absorbed power law plus black body model, scripting the automatic fit for each spectrum in \textsc{xspec} \citep{Arnaud1996}. We assumed interstellar abundances from \citet{Wilms2000} and cross-sections from \citet{Verner1996}. We used the \texttt{tbnew}, \texttt{pegpwrlw}, and \texttt{bbodyrad} models of \textsc{xspec}. In all fits, the neutral hydrogen absorption column density was fixed to $N_{\textsc{H}} = 0.825\times10^{22}$ cm$^{-2}$. We have calculated this quantity considering a variety of relations from the literature. We applied the linear relations between N$_{H}$ and A$_{V}$ given by different authors. We calculated the value of the optical extinction A$_{V}=R_{V}$*E(B-V)=3.7\,mag, from the value of E(B-V)= 1.20\,mag obtained in Sect.\,\ref{sub:redd} and considering $R_{V}$=3.1. From the relation obtained by \citet{Guver2009}, we got $N_{\textsc{H}} = 0.822\times10^{22}$ cm$^{-2}$. A value of $N_{\textsc{H}} = 0.818\times10^{22}$ cm$^{-2}$ is obtained when using the relation given by \citet{Watson2011}. A slightly higher value of $N_{\textsc{H}} = 1.07\times10^{22}$ cm$^{-2}$ is obtained if we use the relation given by \citet{Foight2016}. On the other hand, the relation given by \citet{Willingale2013} from the value of $N_{\textsc{HI}}=0.726\times10^{22}$ cm$^{-2}$ (got with nh of FTOOLS) would provide an estimation of $N_{\textsc{H}} = 0.87\times10^{22}$ cm$^{-2}$. While more complex spectral models have been used in modelling the X-ray spectra of other observatories (see e.g. \citealt{zhang19, jaisawal19}), this approach suffices for an accurate flux determination in the \textit{Swift} band. From the \textit{Swift}/XRT spectral fit, we obtained the fluxes in 0.5--2\,keV and 2--10\,keV bands that will be used to calculate the correlation indexes in Sect.\,\ref{sub:corr}. 

To obtain bolometric fluxes from the \textit{Swift} monitoring, however, the spectral cutoff at higher energies needs to be corrected. Therefore, we combined the above \textit{Swift}/XRT fits with the measured 15--50\,keV count rates from the \textit{Swift}/BAT hard X-ray transient monitor. As the spectral shape changes during the outburst, we cannot simply apply a single, constant conversion factor between the BAT count rate and flux. Instead, we took the best-fit XRT spectral model for each observing date and then added a high-energy exponential cutoff (\texttt{highecut} model in \textsc{xspec}). So that the total model has the functional form of: 
\[
F(E) = \exp(-\sigma N_\textsc{H}) \times \left(\frac{K\times1.0344\times10^{-3}E^{2}}{exp(E/kT)-1} +A E^{-\alpha} \times M(E)\right)
\]

\noindent with $M(E) = \exp\left[\frac{E_\text{c} - E}{E_\text{f}}\right]$ for $E \geq E_\text{c}$ and $M(E)=1.0$ for $E<E_\text{c}$.
Here $K$ is the normalization of the blackbody component, $A$ is the normalization of the power law, and $\alpha$ is the photon index of the power law. The relevant parameters for the high energy cutoff are $E_\text{c}$, the cutoff energy, and $E_f$, the folding energy, both in keV. At energies below $E_\text{c}$, the model is equivalent to the absorbed power law plus blackbody spectrum used for the XRT fits.
We assumed that the exponential cutoff only starts at 8\,keV (i.e., we fixed $E_\text{c}=8$\,keV), with a variable characteristic roll-over or folding energy $E_\text{F}$. This way, any change in the folding energy that is necessary to describe the BAT spectrum, does not influence the XRT data, as they were not used above 8\,keV. At each attempted folding energy, we calculated the corresponding BAT count rate and compared this value to the measured value on that day. Storing the folding energy yielding the best agreement between cutoff-model and observations, we finally calculated the bolometric ($0.1-500$ keV) X-ray flux using the modified spectral model. From this fit, we also obtained the fluxes in the 15--50\,keV band that will be used in Sect.\,\ref{sub:corr}.

\begin{figure}[h]
\begin{center}
\resizebox{\hsize}{!}{\includegraphics{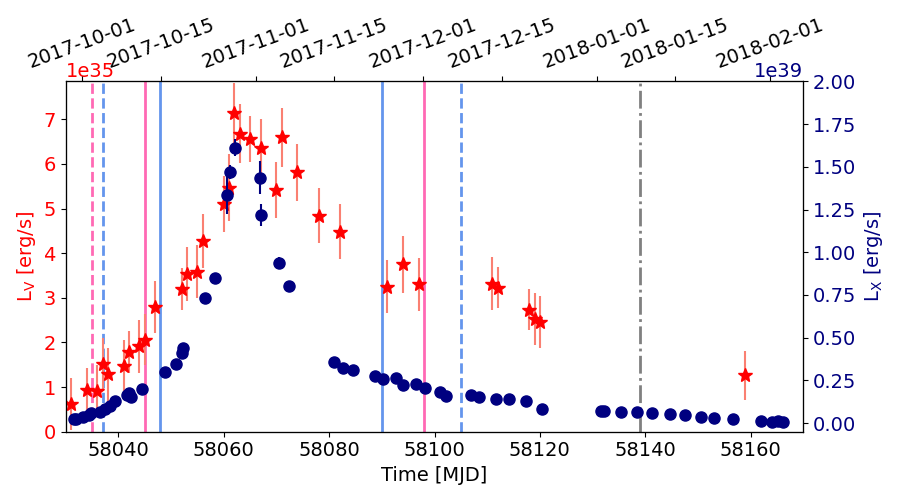}}
\caption[]{Optical vs X-ray bolometric (15--50\,keV) emitted luminosities during the bright outburst of Swift~J0243.6+6124. The optical $V$--band integrated fluxes have been obtained multiplying the monochromatic fluxes (converting magnitudes to fluxes with a zeropoint ZP=3.631$\times 10^{-9}$ erg s$^{-1}$ cm$^-2$ $\AA^{-1}$) by the effective width of the filter (890\AA) and dereddened considering E(B-V)=1.20\,mag A distance d=5.2\,kpc has been assumed in the luminosity calculations ($L=4\pi d^{2}\times F$). The times related to changes in the pulse profiles and the X-ray spectrum proposed by \citet{Liu2022} and \citet{Doroshenko2020} are marked with blue and pink lines respectively. The gray dashed-dotted line was proposed as a critical time (MJD~58139) on both works (see Sect.\,\ref{sub:hump} for details).}
\label{fig:LoptLxLCs}
\end{center}
\end{figure}

The light curves of the residual optical emission in the $V$--Johnson band obtained from the calculations described in Sect.\,\ref{sub:detrend}, along with the X-ray light curve of the Swift~J0243.6+6124 outburst, converted to luminosities considering a distance of d=5.2\,kpc, are shown in Fig.\,\ref{fig:LoptLxLCs}. A clear correlation between the optical and X-ray fluxes is observed, although this correlation differs between the rise and the decay of the outburst.

\subsection{Timescales of the light curves decays}
\label{sub:taus}
As mentioned in Sect.\,\ref{sec:int}, when reprocessing dominates the optical emission during X-ray outbursts of X-ray binaries, the exponential decay timescales of the optical and IR light curves should be 2--4 times longer than the X-ray value \citep{King1998, Rykoff2007}. An exponential decay can be observed for \swift~in Fig.~\ref{fig:LoptLxLCs}.

We have calculated the decay of the outburst in our X-ray, UVOT-$uw2$, UVOT-$um2$,UVOT-$uw1$, and $V$--Johnson light curves, by fitting the formula $F=F_{0}e^{-t/\tau}$ to our observations, and the results of these fits are shown in Table\,\ref{tab:taus}. We could not fit the UVOT-$u$ and UVOT-$b$ light curves due to the poor temporal coverage of the outburst decay that did not allow to correctly perform the fit.

\begin{table}[h]
\caption{Timescales of the decay of the giant outburst for the X-ray, UVOT-$uw2$,UVOT-$um2$,UVOT-$uw1$, and $V$--Johnson light curves, as a result of fitting the formula $F=F_{0}e^{-t/\tau}$ to the observed light curves.}              
\label{tab:taus}      
\begin{center}
\begin{tabular}{ c c c c c}          
\hline\hline                        
  $\tau_{0.1-500\,keV}$ & $\tau_{UW2}$ & $\tau_{UM2}$ & $\tau_{UW1}$ & $\tau_{V}$ \\    
\hline                                   
13$\pm$1 &    22$\pm$3 & 21$\pm$3 & 40$\pm$3 & 42$\pm$3 \\      

\hline                                             
\end{tabular}
\end{center}

\end{table}

The decay timescales of the optical ($V$--Johnson band) and UVOT-$uw1$ light curves are $\sim$4 times the decay timescale of the X-ray light curve. However, the decay timescales of the UVOT-$uw2$ and UVOT-$uw1$ light curves are $\sim$2 times the decay timescale of the X-ray light curve.

\subsection{Flux-flux correlations}
\label{sub:corr}

As mentioned in Sect.\,\ref{sec:int}, a power law correlation $F_{\rm V} \sim F_{X}^{0.5}$ have usually been interpreted as reprocessing of X-rays at longer wavelengths in X-ray binaries \citep{vanParadijs1994, Russell2006}.
\begin{table*}
\caption{Slopes of the correlations between the optical/UV fluxes and the X-ray fluxes in different bands, F$_{\rm opt/UV}$ $\propto$ F$_{X}^{\beta}$, during the outburst.}              
\label{tab:corrFoptFX}      
\begin{center}
\begin{tabular}{l l c c c c c}          
\hline\hline                        
Filter & Epoch & $\beta_{0.1-500\,keV}$ & $\beta_{0.5-2\,keV}$ & $\beta_{2-10\,keV}$ & $\beta_{15-50\,keV}$ & $\beta_{10-300\,keV}$ \\    
\hline                                   
$V$--Johnson & Rise & 0.56$\pm$0.03 & 0.40$\pm$0.02 & 0.47$\pm$0.02 & 0.90$\pm$0.07 & 0.94$\pm$0.08\\      
$V$--Johnson & Decay & 0.37$\pm$0.03 & 0.26$\pm$0.02 & 0.31$\pm$0.02   & 0.59$\pm$0.04 & 0.57$\pm$0.04\\

UVOT-$b$ & Rise & 0.47$\pm$0.04 & 0.35$\pm$0.03 & 0.40$\pm$0.03 & 0.73$\pm$0.05   & 0.75$\pm$0.04 \\
UVOT-$u$ & Rise & 0.49$\pm$0.05 & 0.37$\pm$0.03 & 0.43$\pm$0.04 & 0.7$\pm$0.1   & 0.7$\pm$0.1 \\
UVOT-$uw1$ & Rise & 0.45$\pm$0.03 & 0.34$\pm$0.02 & 0.39$\pm$0.03 & 0.69$\pm$0.07   & 0.69$\pm$0.07 \\
UVOT-$uw1$ & Decay & 0.36$\pm$0.07 & 0.24$\pm$0.05 & 0.30$\pm$0.06 & 0.6$\pm$0.1   & 0.6$\pm$0.1 \\

UVOT-$um2$ & Global & 0.43$\pm$0.07 & 0.31$\pm$0.05 & 0.37$\pm$0.06 & 0.69$\pm$0.11   & 0.70$\pm$0.11 \\
UVOT-$uw2$ & Global & 0.40$\pm$0.04 & 0.28$\pm$0.03 & 0.35$\pm$0.03 & 0.65$\pm$0.06   & 0.66$\pm$0.06 \\

\hline                                             
\end{tabular}
\end{center}

\end{table*}

\begin{figure*}[!]
\begin{center}
\includegraphics[width=0.7\textwidth]{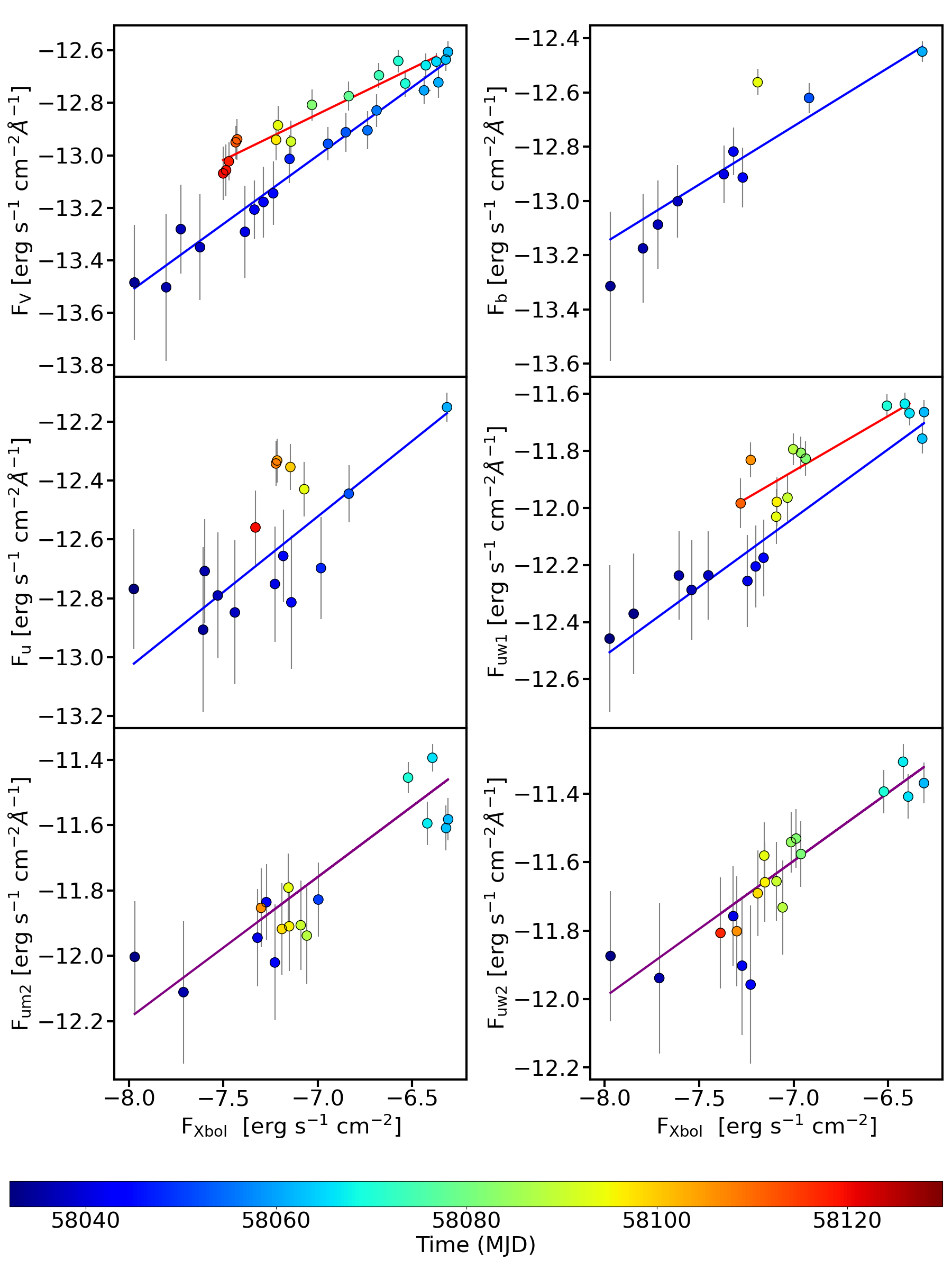}
\caption[]{Optical/UV vs X-ray (0.1--500\,keV) logarithmic fluxes along the outburst of Swift~J0243.6+6124. The correlations show different slopes during the outburst rise than during the decay epochs for the filters with longer wavelengths ($V$--Johnson to UVOT-$uw1$). In this case, different fits have been performed for the rise (blue line) than for the decay (red line). For the UVOT-$um2$ and UVOT-$uw2$ filters, a global fit (purple line) has been performed.}
\label{fig:fluxcorr}
\end{center}
\end{figure*}
We have calculated the indexes of the correlation between the fluxes in each optical/UV filter and simultaneous X-ray fluxes, considering different X-ray energy bands. The correlations between the optical/UV logarithmic fluxes in the different filters and the bolometric logarithmic X-ray fluxes (0.1--500~keV) are represented in Fig.~\ref{fig:fluxcorr}. For the fluxes in the redder wavelengths ($V$--Johnson until UVOT-$uw1$ filters) a different correlation is observed during the rise than during the decay of the outburst. For the $V$--Johnson and UVOT-$uw1$ filters, two fits (for the rise and for the decay) have been performed. Although a different trend is also observed during the rise than during the decay for the observations in the UVOT-$b$ and UVOT-$u$ filters, the temporal coverage for the decay of the outburst was not good enough to perform a reliable fit, and only the fits during the rise of the outburst are provided. For the filters at shorter wavelengths (UVOT-$um2$ and UVOT-$uw2$) this effect is not observed and a global fit provides better results. The results of this correlation analysis are provided in Table\,\ref{tab:corrFoptFX}.

Appart from the different trends between the rise and decay for the redder fluxes, we have also observed that the correlation indexes depend on the energy band used to estimate the X-ray flux, increasing when harder X-rays are used. This effect is observed for all the optical/UV filters (see Table\,\ref{tab:corrFoptFX}).

\section{Results and discussion}
\label{sec:disc}

In this section, we will discuss  first the possible mechanisms that can explain the observed UV/optical fluxes, considering the results obtained in Sect.\,\ref{sec:anal}. Then we will model the most relevant effects in order to reproduce the evolution of the SED along the outburst in correlation with the X-ray emission. We will discuss the possible interpretations of the results of the SED fits. Finally, we will provide some possible explanations to the optical/UV humps observed during the decay of the outburst.

\subsection{Phenomenological constraints on the optical/UV origin}

As we mentioned in Sect\,\ref{sec:int}, there are several mechanisms that can contribute to the optical/UV emission and display similar amplitude variability in X-ray binaries.

Firstly, we will consider the mechanisms inherent to the Be star and/or the circumstellar disk. Long-term variability, either due to formation and dissipation of the Be circumstellar disk and/or to changes in the inclination or due to warping and/or precession of this circumstellar disk, typically produce optical variability with amplitudes of tenths of magnitudes \citep{yan12, moritani13,alfonso17}, but the timescales of this circumstellar disk variability are longer than the duration of the X-ray outbursts of BeXBs. Indeed, X-ray outbursts usually take place when the circumstellar disk reaches a critical size, and/or the inclination of the circumstellar disk favours the accretion onto the NS. On the other hand, optical outbursts due to mass ejections from the Be star with durations of several days, and displaying amplitudes of tenths of magnitudes are commonly observed in classical Be stars \citep{Rivinius2013} and in BeXBs \citep{yan12, alfonso17}. However, when ejections of matter from the donor star take place, a delay between the X-ray emission and the optical outburst would be expected. Indeed, they can cause the replenishment of the Be circumstellar disks, and these changes in the Be disk could trigger an X-ray outburst \citep{yan12}. Then, these mechanisms can not explain the tight correlation and strict simultaneity between the optical/UV fluxes and X-ray fluxes that we observe during the giant outburst of \swift.

We also considered that the optical emission could be originated in the jet, given the fact that this was the first Be X-ray binary for which signatures from a jet has been found \citep{vandenEijnden2018}. We have estimated the contribution of the optical emission considering the radio spectral shape, which is steep at the peak of the outburst ($\alpha$<0). If we take the brightest observed radio flux at 58073~MJD, which was 92\,$\mu$Jy at 6\,GHz, an we use the spectral index measured by \citet{vandenEijnden2018} $\alpha$=-0.64, we obtain an extrapolated optical flux of $F_{V}\sim 6.1\times 10^{-20}$ \esca. If we do the same calculation for the observations at 58090~MJD, when the slope was flatter, $\alpha=0.08$, we obtain an optical contribution from the jet of $F_{V}\sim 6.2\times 10^{-17}$ \esca, still very small compared to the observed variability intrinsic to the outburst of $\Delta F_{\rm V}\sim 2.3\times10^{-13}$ \esca (see Fig.\,\ref{fig:fluxcorr}). We can then rule out the emission coming from the jet as the main origin of the optical/UV emission. Moreover, there were also observations after the optical outburst with similar radio fluxes when the optical fluxes were fainter, so that we can conclude that the optical and radio emission are not correlated in this system. We want to point out that the BeXB radio jets are really faint: $\sim$6200 times fainter than in black hole X-ray binaries (see \citealt{vandenEijnden2022}). That could explain why you may possibly have jet contributions to the OIR emission in BH LMXBs, but not for BeXBs.

There are of course, other mechanisms that could lead to optical/UV emission directly correlated with the X-ray emission and have to be considered in order to explain the optical and UV fluxes observed during this outburst. These mechanisms are: {\em a)} thermal emission from the viscously heated accretion disk around the NS, {\em b)} reprocessing of the hard X-rays in the accretion disk, c) X-ray heating of the surface of the Be star. Given the tight correlation observed between the X-ray and optical/UV fluxes, some of these X-ray related mechanisms could be significantly contributing to the optical and UV emission observed during the giant X-ray outburst of \swift. The UV and optical timescales calculated in Sect.\,\ref{sub:taus} are in the range 2--4 times the X-ray decay timescale (see Table\,\ref{tab:taus}), so according to \citet{King1998}, these results favour the X-ray reprocessing as the main mechanism contributing to the UV and optical variability observed during the outburst. The differences found between the timescales of UVOT-$uw2$ and UVOT-$um2$ and those obtained for UVOT-$uw1$, and $V$--Johnson (the last ones are around twice longer than the other ones), could be indicating that there are differences between the origin of the emission on the different bands. The correlation indexes obtained from the flux-flux correlation analysis ($F_{\rm V} \propto F_{X}^{\beta}$) are shown in Table\,\ref{tab:corrFoptFX}. Some of the obtained indexes,  would be in agreement with the correlation index typical of X-ray reprocessing proposed by \citet{vanParadijs1994}, $\beta$=0.5. However, the obtained correlation indexes vary between the different optical/UV filters and when different X-ray bands are used. Moreover, for the redder observations, these indexes are larger for the rise than for the decay of the outburst. The dependence on the energy band could be related to the changes in the X-ray spectra, and the variable contribution from the reflected emission during the Super-Eddington state \citep{Bykov2022}. Also the fact that the albedo is expected to vary with the energy of the X-ray impinging radiation could affect to the correlations (harder X-rays penetrate more in the disk). The differences between the rise and decay could be related to the presence of a more extended accretion disk during the decay than during the rise, or if the disk gets flared along the outburst. Moreover, the source underwent different accretion regimes, from sub- to super-critical accretion, and to a third regime with a radiation-pressure dominated accretion disk \citep{Doroshenko2020}, and this could also affect to the correlations. As mentioned by \citet{Gierlinski2009}, these simple flux-flux correlations alone can be inaccurate to explain in detail the origin of the optical/UV emission for individual LMXBs. For HMXBs, the analysis can be even more complicated. For this reason, we tried a different approach involving a more complex modelling of the emission mechanisms that will be presented in Sect.\,\ref{subsec:SEDmod}.

We also wanted to give a possible explanation to the singularity of this optical/UV outburst. The unprecedent high X-ray luminosities for a BeXB observed during this giant outburst, being the first Ultraluminous X-ray Pulsar (ULXP) identified in the Galaxy, could have led to the exceptionally bright outburst at optical and UV wavelengths. X-ray outbursts of BeXBs typically have luminosities of $L_{X}$=10$^{36}$--10$^{37}$\,erg/s. For the orbital parameters of \swift, and considering the intrinsic brightness of the optical companion, an X-ray outburst of $L_{X}$=10$^{37}$erg/s would produce an observed optical outburst with amplitude $\Delta$V=0.06\,mag, and an X-ray outburst of $L_{X}$=10$^{36}$erg/s would lead to an optical outburst with $\Delta$V=0.02\,mag.   These values are close to typical uncertainties in optical photometric measurements in many cases, and are much smaller than the amplitude we observe in our optical outburst. 
On the other hand, the orbital period of \swift\,(P$_{orb}$=27.587\,days), and hence the distance between the NS and the Be star, is shorter than for other BeXBs that have underwent very bright giant outbursts, such as 1A 0535+262, which reached L$_{X}\sim$ 1 $\times$10$^{38}$erg s$^{-1}$ \citep{Reig2022}, and whose orbital period is P$_{orb}$=111.1\,d \citep{Finger2006}.
Then, we can explain the exceptionality of this optical/UV outbursts as a combination of the extreme X-ray luminosities, together with a semimajor axis value shorter than for other BeXB systems.

\subsection{SED modelling through the outburst}
\label{subsec:SEDmod}
Since the effects directly related to the X-ray emission seem to be the most plausible mechanisms driving the optical/UV emission observed during the giant ourburst of \swift, we have created a physical model considering the following mechanisms: emission from gravitational energy released through viscosity, emission coming from the X-ray irradiation of the accretion disk, and emission due to the heating of the surface of the companion star. We have modelled these effects and used the calculated fluxes in each optical/UV filter to build a theoretical SED to be compared with the observed optical/UV SEDs.

\subsubsection{Description of the model}

In all the calculations, we have used a distance of $d = 5.2$\,kpc to estimate the X-ray incident luminosities (which is an input for the three models), and to convert the calculated optical/UV luminosities to fluxes for comparison with the observed SEDs. 

We have used the most recent orbital parameters provided in the GBM Accreting Pulsar Histories web accesible at \url{https://gammaray.nsstc.nasa.gov/gbm/science/pulsars/lightcurves/swiftj0243.html} of the GBM Accreting Pulsars Program (GAPP, \citealt{Malacaria2020}). We then used an orbital period P$_{orb}$= 27.698899\,days, a projected semi-major axis $a\,\sin$i=115.531 light-sec, an eccentricity $e$=0.1029, and a binary epoch T$_{\pi/2}$=2458116.09700 JED. On the other hand, we considered an inclination of the orbit of $i$=30$^{\circ}$ (Paper~I).

\begin{figure}
\begin{center}
\resizebox{\hsize}{!}{\includegraphics[width=1.3\textwidth]{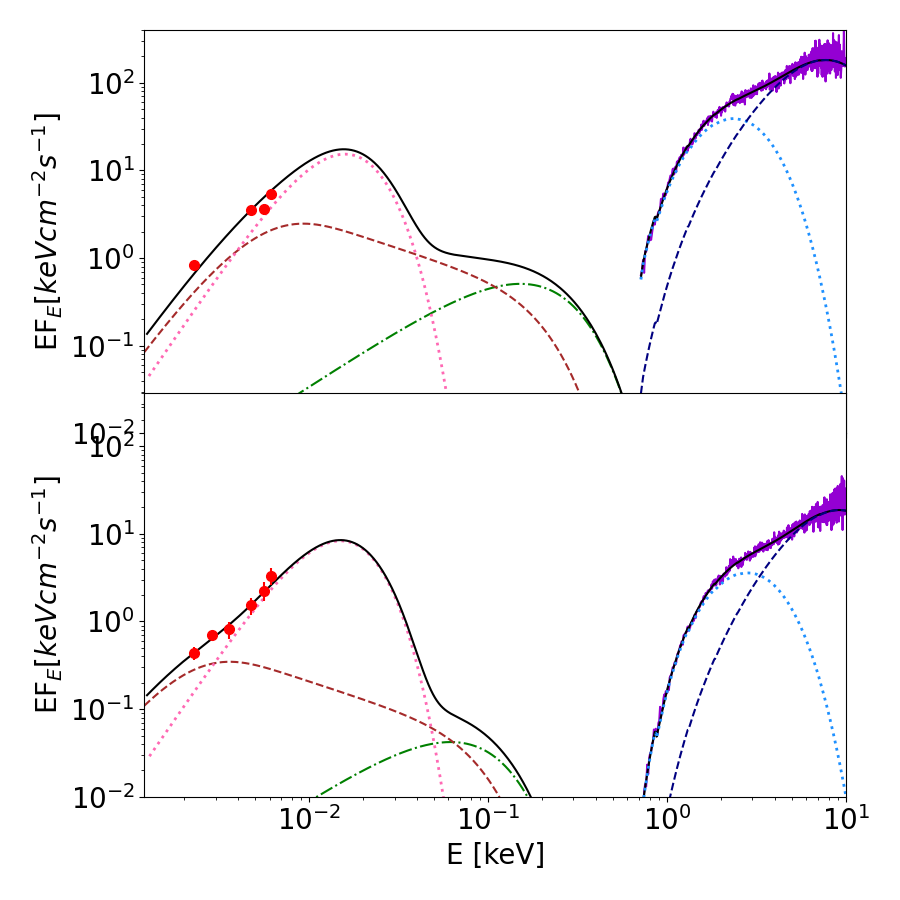}}\\
\caption[]{Models considered in this work to fit the optical/UV observations (red points) and XRT spectra (purple line) for days 58062~MJD (up) and 58094~MJD (bottom). The contribution of X-ray heating of the companion star is represented with a pink dotted line, the irradiated disk is plotted as a brown dashed line, the viscously heated disk is plotted with a green dash-dotted line. The xspec models \texttt{bbodyrad} and \texttt{pegpowerlaw} used to fit the XRT spectra are plotted with light blue dotted and navy blue dashed lines respectively.}
\label{fig:models}
\end{center}
\end{figure}

For each day of the outburst (whenever we had some optical/UV observation), we used the X-ray luminosity extracted from linear interpolation of the X-ray bolometric fluxes in the light curve generated from the method described in Sect.\,\ref{sub:estX}, to be used as an input of the different parts of the model. The XRT spectra (purple line) and the models used to fit them for two representative days of the outburst, one close to the peak of the outburst (58062~MJD, up) and another one during the decay of the outburst (58094~MJD, bottom), are plotted in Fig.\,\ref{fig:models}. The blackbody components are plotted with light blue dotted lines and the power laws are plotted with navy blue dashed lines. 

For the two mechanisms involving the accretion disk, we performed two different fits. In the first one, we left the outer radius $R_{out}$ as a free parameter, using a grid with step 0.01$a$, and taking values from $R_{out}$=0.01$a$ to $R_{out}$=0.20$a$ (typical predicted sizes of an accretion disk according to theoretical models are $R_{out}=0.04-0.10a$, see for example \citet{Hayasaki2004, Hayasaki2006})), being $a$ the semi-major axis of the orbit. We have set an upper limit of $R_{out}$ based on the size of the Roche lobe. For a mass of the donor star of M$\ast$=18\,$M\sun$ and considering the mass of the NS M$_{NS}$=1.4\,$M\sun$, and using the approximation by \citet{Eggleton1983}, we got a value of the Roche lobe of $R_{L}$=0.20$a$. For the second fit, we fixed $R_{out}=0.20a$ and fit a factor dependent on other effects that will be explained later. For each accretion rate, we approximated the inner radius of the accretion disk as the magnetospheric radius, calculated considering a magnetic field of B$\sim$6.6*10$^{12}$G (value obtained from the estimation by \citet{Wilson-Hodge2018}, B$\sim 1\times10^{13}$\,(d/7kpc)\,G). To estimate $r_{m}$ we used the following equation \citep{Furst2017}:

\begin{equation}
 r_{m}=\dot{M}^{-2/7}\mu^{4/7}(GM)^{-1/7}2^{-3/7}
\end{equation}

The accretion rate ($\dot{M}$) has been obtained from the X-ray bolometric luminosity using the formula:

\begin{equation}
    L_{X}=\frac{GM\dot{M}}{R}
\end{equation}
and considering typical values of a NS of $M = 1.4 \msun$\,and $R = 10^{6}$\,cm.

We describe below the details of the specific models we considered for each of these three mechanisms. These components of the model used to fit the optical/UV observations (red points), are plotted in Fig.\,\ref{fig:models}.

\begin{itemize}
    \item Emission from the viscously heated accretion disk:\\

Intrinsic thermal emission from the viscously heated outer accretion disk can contribute to the X-ray, UV and optical emission of X-ray binaries \citep{Shakura1973, Frank2002}. 

We modelled the contribution of a viscously heated accretion disk, using the approximation of an optically thick disk \citep{Shakura1973}, using the formula:

\begin{equation}
 T(R)=\left\lbrace \frac{3GM\dot{M}}{8\pi R^{3} \sigma} \left[ 1-\left( \frac{R_{*}}{R} \right) ^{1/2} \right] \right\rbrace ^{1/4}
\end{equation} 

We calculated the temperature for each radius between R$_{in}$ and R$_{out}$, and integrated the Planck law over each area element of the disk to estimate the emitted spectrum. 

The modelled emission due to this effect  is plotted with green dash-dotted lines in Fig.\,\ref{fig:models}. As can be seen, due to the effect of the magnetic field on the inner radius, this emission does not contribute to the soft X-rays in the energy range covered by the XRT observations (purple line).  \\

\item X-ray irradiation of the accretion disk:\\

We have modelled the effect of the X-ray irradiation of the accretion disk for each incident $L_{X}$ along the outburst, for the grid of values of $R_{out}$ described above. We have modelled the emission from an irradiated disk using the calculations in \citet{Vrtilek1990}, which give this relation for T(R): 

\begin{equation}
 T(R) \simeq 23200K \left(\frac{f_{2}f_{3}\sqrt{f_{1}}}{0.5}\right)^{2/7} \left(\frac{\dot{M}}{10^{18}g/s}\right)^{2/7} \left(\frac{\rsun}{R}\right)^{3/7}
 \label{eq:irrd}
\end{equation}

where $f_{1}$ is a factor that depends on the details of the vertical structure of the disk and has been fixed to 1, $f_{2}$ is the absorbed fraction of the impinging radiation  and has been fixed to 0.5, and $f_{3}$ takes into account the possible anisotropy of the radiation and has been fixed to 1.

The modelled emission due to this effect  is represented  with brown dashed lines in Fig.\,\ref{fig:models}.\\

\item X-ray heating of the surface of the companion star:\\

The irradiation of the surface of the the Be star by the X-ray photons emitted close to the NS can contribute significantly to the optical and UV emission \citep{vanParadijs1995, Charles2006}. We have calculated the extra emission in each band by estimating the difference between the emission from an unirradiated O9.5\,V star and that from an irradiated star, considering a radius of the donor star of $R_{\ast}=8R_{\sun}$, and an initial effective temperature of T$_{\mathrm{eff}}$=32000\,K. 

We considered that the X-rays emitted by the NS are absorbed by the atmosphere of the optical companion and then thermally re-radiated at lower energies. We have followed a similar modelling to the one describe in \citet{ducci19} to estimate the contribution of the heated surface of the Be star on A0538--66. In this model, the X-rays emitted by the central source are absorbed by the atmosphere of the companion star and heat its surface. For each surface element of the star, the local effective temperature increases as:

\begin{equation}
 T^{4}_{e} = T^{4}_{0,e}+(1-\eta)\frac{L_{X} cos \phi}{4 \pi \sigma R^{2}}
 \label{eq:hs}
\end{equation}

with $T^{4}_{0,e}$ the effective temperature of an unirradiated surface element, $\eta$ is the albedo (fraction of reflected X-rays) that we have fixed to 0.5, $L_{X}$ is the X-ray luminosity of the source, $\phi$ is the angle between the direction to the X-ray source and the normal of the surface element, $\sigma$ is the Stefan-Boltzmann constant, and $R$ is the distance from the surface element to the X-ray source. We calculated this distance for each phase using the latest orbital parameters.

\begin{figure}
\begin{center}
\resizebox{\hsize}{!}{\includegraphics[width=\textwidth, height=0.7\textheight]{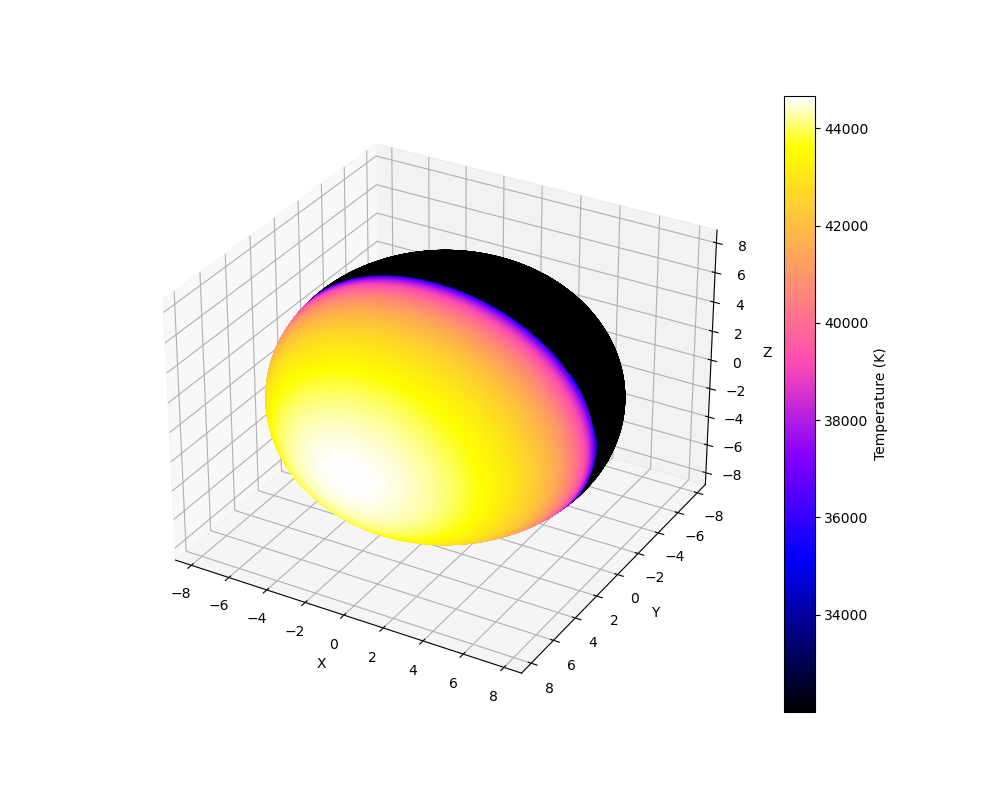}}\\
\caption[]{Illustration of the effect of X-ray heating  of the surface of the companion star. The calculations for the peak of the outburst (58064\,MJD) are shown.}
\label{fig:heated_star}
\end{center}
\end{figure}

We have calculated the contribution of this effect for each X-ray luminosity along the outburst. To illustrate this effect, the modelled heating of the surface of the optical companion of \swift\, at the peak of the outburst (MJD\,58064) is shown in Fig.\,\ref{fig:heated_star}. As can be seen, the center of the facing star reaches an extra heating of $\Delta T>12000$\,K.

We have tried to simplify this contribution of the model as much as possible and we are aware that there are some considerations that we have not taken into account. We want to point out that, since the temperatures would be always larger than the effective temperature of the star, the emitted spectrum would always peak at wavelengths shorter than 905\,\AA\,($\lambda_{\textrm peak}$ for T$_{\textrm eff}$=32000\,K from Wien's law). This corresponds to energies beyond those in the range of our optical/UV observations, which all fall on the Rayleigh-Jeans tail of this emission (see Fig.\,\ref{fig:models}, the emission reprocessed by the Be star is plotted as a pink dotted line). This implies that  systematic uncertainties on the details of the reprocessing process would mainly have a small impact on the normalization only (increasing or decreasing the modelled continuum emission), but not on the shape of the spectrum itself. 

One example of such an uncertainty is the exact value of the albedo, which affects the range of temperatures reached by the heated stellar atmosphere (see Eq.\,\ref{eq:hs}). We have made several tests modifying this value and have concluded that, compared to the model represented in Fig.\,\ref{fig:models}, these modifications would cause only small changes in the normalization, but not in the spectral shape at the observed wavelengths. 

Another uncertainty is the role of the stellar rotation: Eq.\,\ref{eq:hs} implicitly assumes that the stellar atmosphere is able to reach thermal equilibrium with the incoming X-ray flux on timescales significantly shorter than the rotational timescale of the star. In Swift J0243.6+6124, that timescale is of the order $\tau_{\textrm rot}\sim 1$ day, if we assume the values of the projected rotational velocity $vsini$=210\,km/s and inclination $i$=30$^{0}$ from \citet{Reig2020}. From the correlation observed between the optical and X-ray observed fluxes, we can conclude that the cooling/heating timescale should be shorter than the rotational period. 

But if the timescale was of the order or longer than the rotational period, the X-rays would produce a temperature profile with constant-temperature rings at each latitude,  and with peak temperatures slightly lower than when no rotation is considered (due to the redistribution of the incoming energy). In any case, both faces of the star would be heated, although they would reach maximum peak temperatures slightly lower (e.g. $\sim$40700~K vs. 44700~K in the equator of the Be star at the peak of the outburst), and the modelled emission would decrease by a factor of $\sim$0.8 in the spectral range of our observations. 

On the other hand, we are not considering that some fraction of the donor star could be obscured by the circumstellar disk, although the height scale is expected to be small \citep{Okazaki2002}.

Modifying the fraction of the surface of the Be star that is heated would also lead to a variation in the normalization of the blackbody emission. Detailed calculations of these timescales, as well as extra effects such as differential rotation and convection that further alter the temperature profile, or modelling the occultation by the circumstellar disk would introduce a large set of new assumptions, compared to the limited number of fitted datapoints. We therefore leave the consideration of these effects to future modelling work with more complete datasets.

\end{itemize}

We also want to mention that we have not included in our modelling the effect of the irradiation of the Be circumstellar disk. It should be taken into account that there are some important differences between circumstellar disks in BeXBs and Be disks in classical isolated Be stars. The correlation between the orbital period in BeXBs and the H$_{\alpha}$ equivalent width has been interpreted as evidence of the truncation of the Be disk \citep{Reig2011}. These disks can be truncated at different resonant radii, where the angular frequency of the disk and that of the orbit is an integer \citep{Monageng2017}, and can reach the size of the Roche lobe during X-ray outbursts (which is R$_{L}\sim$10\,R$\ast$ for the donor star in this system). However, due to the resonant torque the outer disc density drops much more rapidly in Be X-ray binaries than in isolated Be stars \citep{Okazaki2002}. 

In our modelling, we assumed that the Be disk was coplanar to the orbit during the outburst, which is the simplest configuration to explain the occurrence of an X-ray outburst in a BeXB system \citep{Okazaki2002}. We are aware that some proposed scenarios to explain the occurrence of giant outbursts involve different configurations, including precession of warped Be disks \citep{martin2011} and eccentric and tilted disks due to Kozai-Lidov oscillations \citep{martin14a, Martin2019}. Indeed, for some BeXBs, signatures of tidal warped circumstellar disks have been observed during giant X-ray outbursts \citep{moritani13, Reig2018}. However, no signs of warping and/or eccentricity have been found in the profiles of the optical emission lines during the giant outburst of Swift J0243.6+6124 \citep{Reig2020,Liu2022b}, although we can not rule out the presence of a tilted disk with the tilting axis almost perpendicular to the line of sight (in this case the emission lines would also be single-peaked).

Under the considered assumption that the Be disk is co-planar to the orbit, the X-rays would not impinge on the Be disk, since the angles that the X-rays form with the normal vector of the Be disk surface would be close to 90 degrees and then they would not heat the Be disk surface. Moreover, a significant fraction of the X-rays inciding with  such small angles w.r.t. the orbital plane could be firstly intercepted by the outer parts of the flared accretion disk. Then, under this assumption, only the outer, low-density, and colder material of the disk could be heated. The emission of this colder material, if irradiated, would probably peak at redder wavelengths. We consider that correctly modelling the temperature and density gradient and the geometry of the Be disk, and the effect of the X-ray irradiation of this disk would require many more assumptions, and that performing such a detailed analysis with the dataset we present in this work is out of the scope of this paper. However, we want to point out that not including this contribution could lead to an overestimation of the outer radius of the accretion disk (see Sect.\,\ref{sub:SEDfits}), since similar temperatures could be reached in the outer regions of both disks.

\subsubsection{Optical/UV SEDs fits}
\label{sub:SEDfits}

\begin{figure*}
\begin{center}
\includegraphics[width=0.96\textwidth]{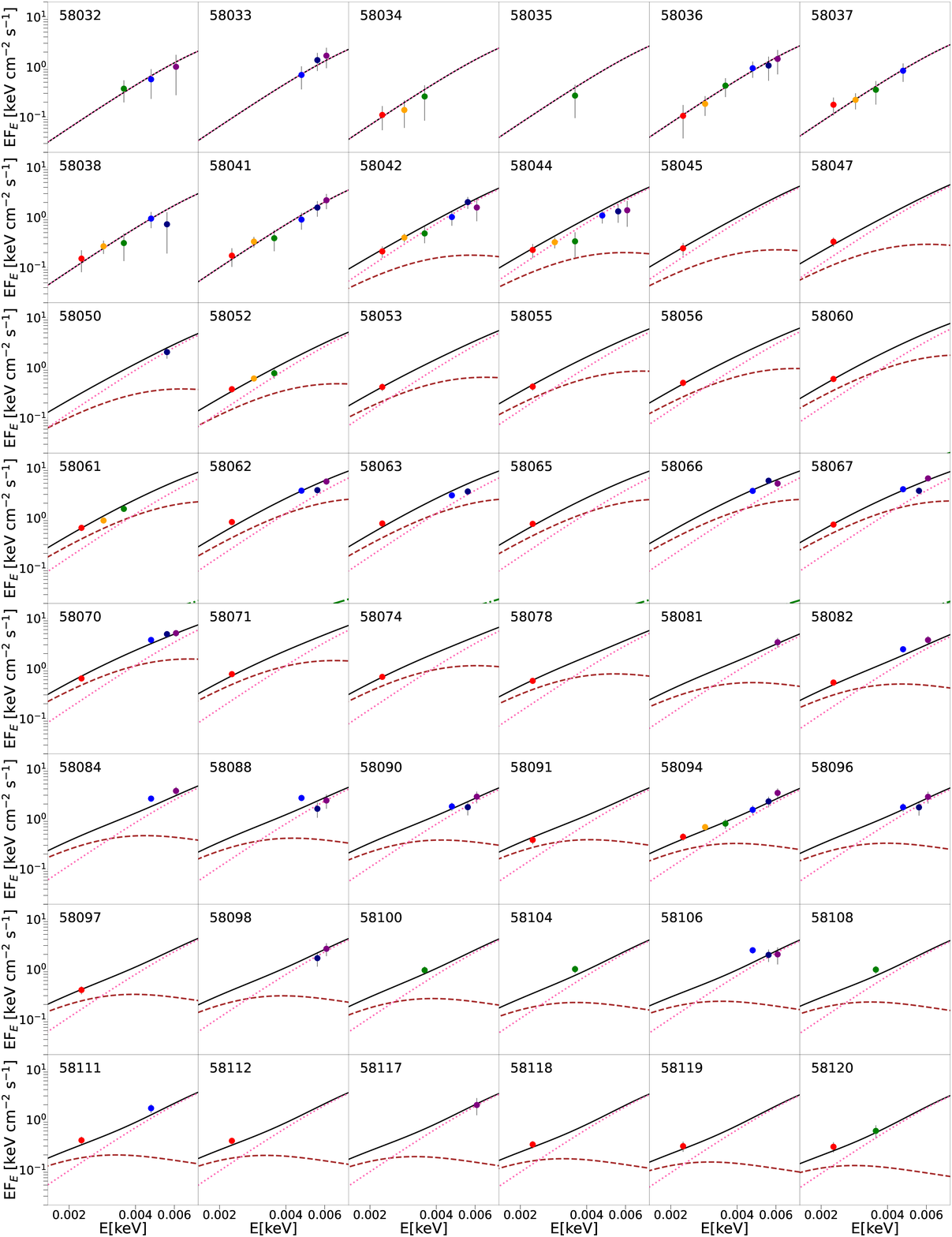}
\caption[]{Optical/UV SEDs evolution along the outburst. The reddening corrected observed fluxes are plotted with different colours: $V$-Johnson in red, UVOT-$b$ in yellow, UVOT-$u$ in green, UVOT-$uw1$ in blue, UVOT-$um2$ in navy blue, and UVOT-$uw2$ in purple. Error bars are plotted in gray. Best fitted model predictions are plotted for comparison (black solid lines). The pink dotted lines represent the modelled contribution by the extra emission from heating the surface of the companion star. The brown dashed lines correspond to the contribution from an irradiated accretion disk with the value of $R_{out}$ providing the best fit.}
\label{fig:SEDs}
\end{center}
\end{figure*}

\begin{figure}
\begin{center}
\includegraphics[width=0.5\textwidth]{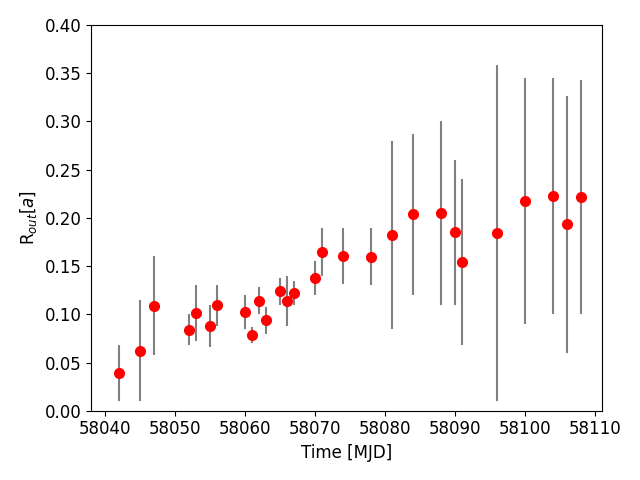}
\caption[]{Outer radius of the accretion disk which provided the best fit for each SED.}
\label{fig:Routs}
\end{center}
\end{figure}

The optical/UV SEDs built with the available observations along the outburst, together with the best SED fit for each day are plotted in Fig.\,\ref{fig:SEDs}. For those days for which we had optical measurements in the $V$ band, we calculated the value of the outer radius of the accretion disk, $R_{out}$, yielding the best fit of the SED using the minimization of chi-square method (see Fig.\,\ref{fig:Routs}). The black solid lines represent the sum of the three modelled components. The different mechanisms are plotted with different colours and line styles: the pink dotted lines represent the modelled contribution by the extra emission from heating the surface of the companion star, and the brown dashed lines correspond to the emission modelled from an irradiated accretion disk with the value of $R_{out}$ providing the best fit. The emission from the viscously heated accretion disk is too faint to appear represented in the plot. Since it does not contribute significantly to the optical and UV fluxes and can be ruled out from the discussion.

As can be seen in Fig.\,\ref{fig:SEDs}, the heating of the surface of the companion star (blue dotted line) contributes significantly to the optical and UV observed fluxes along the whole outburst. The SED fits get improved with the addition of an irradiated disk from MJD~58047 to 58120. Before MJD~58047, and from MJD~58141 onwards, the heating of the companion star is enough to explain the observed optical/UV emission. Although from MJD~58141 onwards, there were not enough optical observations to fit the redder part of the SED (more affected by irradiation of the accretion disk at these X-ray luminosities). The irradiation of the accretion disk contributes significantly to the UV emission close to the peak of the outburst (from MJD 58062 to 58096 aprox.).

As mentioned above, to reproduce the observed fluxes, we performed two different fits. First, we left the outer radius of the accretion disk ($R_{out}$) as a free parameter of the fit (see Fig.\,\ref{fig:Routs}). From the results of the fits, an increasing $R_{out}$ along the outburst is obtained, reaching a value of $R_{out}\sim$0.18$a$ at 58097\,MJD. After that, a larger value of $R_{out}$, reaching the upper limit given by the Roche Lobe $R_{out}\sim R_{L}=0.20a$, is required to fit the observed optical fluxes from 58100 to 58120\,MJD (see Fig.\,\ref{fig:Routs}). This could be understood as an increase of the size of $R_{out}$ and could be interpreted as a progressive increase of the size of the accretion disk or a dissipation towards the outer parts of the disk. We want to stress that for this first attempt, we fixed all the geometrical factors and the fraction of reprocessed emission along the whole outburst, and we only considered  $R_{out}$ as a free parameter in our modelling. This increase of the emission contributing to the redder wavelengths along the outburst is in agreement with the differences observed between the rise and decay in the flux-flux correlations for the redder wavelengths (Sect.\,\ref{sub:corr}).

\begin{figure}[!]
\begin{center}
\includegraphics[width=0.5\textwidth]{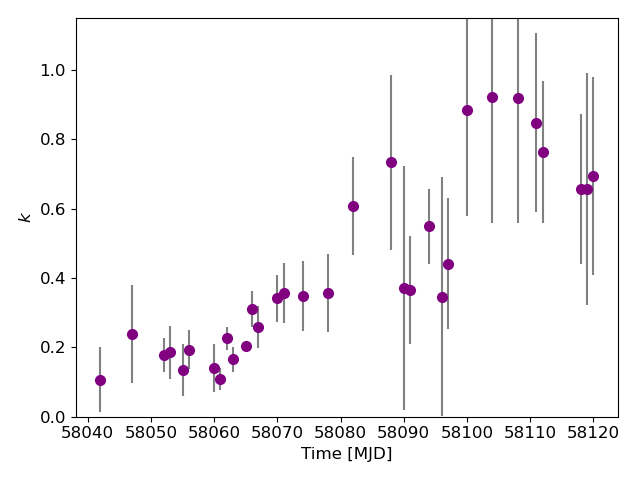}
\caption[]{Results of the SED fit when the outer radius of the accretion disk is fixed to $R_{out}$=0.2$a$, and we use as free parameter the factor correlated with the geometry and fraction of the radiation reprocessed ($k$).}
\label{fig:kgeos}
\end{center}
\end{figure}

Such an increase on the redder fluxes could also be reproduced by modifying the factors in the irradiation disk model which depend on the vertical scale of the disk and the effects of possible anisotropy of the radiation, or the fraction of reprocessed emission due to changes in the ionization or the spectra shape (see Eq.\,\ref{eq:irrd}). Such effects seem to have occurred during this outburst. \citet{Doroshenko2020} proposed that during this giant outburst, the system underwent transitions between three states: sub-critical gas pressure dominated disk (GPD), super-critical GPD (dashed blue line in Fig.\,\ref{fig:LoptLxLCs}), and super-critical radiation pressure dominated disk (RPD). The times proposed for the transitions between these states are marked as dashed and solid pink lines respectively in Fig.\,\ref{fig:LoptLxLCs}. These changes in the geometry of the accretion disk should have an effect on the fraction of reprocessed X-ray emission at longer wavelengths. Moreover, the change from pencil to fan beam regime proposed by \citet{Liu2022} also should have a direct impact on the fraction of X-ray flux impinging on the accretion disk due to geometrical effects. It could also happen that the accretion disk gets flared, and this would result in a stronger heating of the outer parts of the disk and a larger contribution to the redder fluxes. Moreover, the X-ray radiation was harder during the decay than during the rise of the outburst (for similar X-ray fluxes), and the albedo is smaller for harder X-rays. A smaller albedo would produce higher optical/UV emission. To quantify these effects, we have also performed the SED fits fixing $R_{out}$=0.20$a$, and leaving as free parameter a combined factor $k=f_{2}f_{3}\sqrt{f_{1}}$, which includes the three factors present in Eq.\,\,\ref{eq:hs}. The results of these fits are shown in Fig.\,\ref{fig:kgeos}. As expected, and increase of this value is obtained along the outburst.

On the other hand, it is believed that the accretion disk displayed outflows during the brightest part of the outburst \citep{vandenEijnden2019b}, and an extra contribution to the optical continuum emission could be originated in the base of this disk wind \citep{Koljonen2023}. Finally, the timescales of the cooling of the accretion disk and the donor star could have also an effect on the observed 
evolution along the outburst.

\subsection{Optical hump at around MJD 58095--58112 and end of the outburst}
\label{sub:hump}
If we look in detail to the optical/UV light curves of \swift, a hump from around MJD~58095 to 58112 is observed during the decay of the outburst at all the optical wavelengths and also in the UVOT-$uw1$ filter (see Fig.~\ref{fig:optUVLCs}). 

In the previous section, we mentioned that the observed increase on the contribution of the irradiated accretion disk along the outburst could be related with changes in the geometry. \citet{Liu2022} studied the variability of the X-ray pulse profiles along the outburst and identified a high-double peaked pulse profile present from MJD~58048 to 58090, and called it the accretion column regime (see blue solid lines in Fig.\,\ref{fig:LoptLxLCs}). \citet{Doroshenko2020} proposed that a transition between the GPD to RPD disk and viceversa took place at around MJD~58045 and 58098 (pink solid lines in Fig.\,\ref{fig:LoptLxLCs}). During this epoch, we observed a progressive increase of the $R_{out}$ and $k$ values obtained from the SED fits (see Fig.\,\ref{fig:Routs}\, and Fig.\,\ref{fig:kgeos}). After MJD~58097, a sudden change in the fitted values is observed. According to \citet{Liu2022}, the pulse-profile gradually changed from double- to single-peaked from MJD~58091  to 58105 when the pulse profile was single peaked again (dashed blue line in Fig.\,\ref{fig:LoptLxLCs}). This transition epoch would be roughly coincident with the epoch when we observe the optical/UV humps and the change in the fitted $R_{out}$ and $k$ is observed. On the other hand, if a transition from RPD to GPD took place at around MJD~58098, it should be related with the plateau of the optical and UV fluxes. A sudden change in the geometry of the inner flow would be also in agreement with the change on the fitted values of the combined factor $k$ at around MJD~58098 (see Fig.\,\ref{fig:kgeos}). If the optically thick curtain vanishes, there should be an increase of the X-ray irradiation impigning the accretion disk that could produce the observed humps from MJD~58095 to 58115 and the sudden increase on the fitted values of  $k$.

From MJD~58105 to 58139, the source transited from the single-peak pulse profile to the low double-peaked profile, that has been interpreted as the transition from accretion column to the low pencil beam regime \citep{Liu2022}. The impinging X-ray radiation onto the accretion disk is expected to be lower for the pencil beam geometry than for the fan beam case and this could explain the change on the decay trend in the optical light curves from MJD~$\sim$58112 onwards (see Fig.\,\ref{fig:LoptLxLCs}). Indeed, the secondary maxima in the pulse profiles starts shifting on phase from MJD~58112 to 58139 pulse profiles (see Fig.\,1 of \citet{Liu2022}). 

\citet{Liu2022} observed that the pulse profile changed to double-peaked again at around MJD~58139 (dashed-dotted gray line in Fig.\,\ref{fig:LoptLxLCs}), and this change has been interpreted as the transition from super- to sub-critical accretion and as the offset of the accretion column and the presence of a pencil beam dominated accretion \citep{Doroshenko2020,Liu2022}. This is in agreement with the fact from MJD~58140 onwards, the optical/UV SEDs can be reproduced just by the heating of the companion star, and the irradiation of an accretion disk is nos necessary to fit the observations.

Another possible explanation to this optical hump could be that a change in the geometry of the Be circumstellar disk had occurred. A change in the inclination and/or in the precession of the Be disk could be responsible for an increase of the optical/UV emission (with a major contribution at redder wavelengths), and could also be related with the end of the outburst.

\section{Summary and conclusions}
\label{sec:sum}

We have studied the optical/UV emission during the 2017--2018 Super-Eddington giant outburst of \swift.
We have considered the possible origins of this emission and concluded that the main mechanism that can explain the observed light curves is thermal reprocessing of the X-ray emission by the donor star and the accretion disc.

We have estimated an interstellar reddening of $E(B-V)$=1.20$\pm$0.01\,mag from fitting the SED corresponding to X-ray quiescence. We have calculated the flux conversion factors to be applied when converting observed magnitudes in the UVOT filters to monochromatic fluxes. We have joined these UVOT fluxes with the compilation of all the $V$--Johnson band observations from \citet{Reig2020} to build the SED for each day of the outburst (whenever there were available observations).

We have studied the decay timescales of the optical and UV light curves which are between 2--4 times the X-ray light curve timescale. We have also calculated the correlation indexes of $F_{\rm opt/UV}$ $\propto F_{X}^{\beta}$ and obtained values of $\beta$ which depends on the considered X-ray band and on the optical/UV filter, and are larger for the rise than for the decay phase of the outburst for the redder wavelengths. This difference between the rise and decay is not observed for the UVOT-$um2$ and UVOT-$uw2$ filters. The correlation indexes are close to $\beta\sim$~0.5 when considering bolometric X-ray luminosity during the rise of the outburst, which is compatible with the classical value expected from reprocessing proposed by \citet{vanParadijs1994}, but is smaller when the soft X-ray energy bands are used and for the decay of the outburst. Then, to be able to reproduce the observed optical/UV fluxes of \swift, a more sophisticated approach is required. 

For this reason we built a physical model including the contribution of the X-ray heating of the companion star surface, the irradiation of the accretion disk, and the contribution from a viscously heated accretion disk. As an input of the model, we estimated the X-ray luminosity from the spectral fits performed to all the available \textit{Swift}/XRT data considering a power law plus a black body model, and then we added a high-E cutoff with variable folding energy to fit the \textit{Swift}/BAT observed count rates. From the SED fits, we obtained that the principal mechanism contributing to the UV fluxes along the whole outburst is the heating of the companion star. We also concluded that the irradiation of the accretion disk needs to be considered in order to reproduce the observed optical fluxes SEDs from MJD~58047 to 58120 and contributes significantly to the UV emission close to the peak of the outburst. Before and after these times, the heating of the companion star is enough to reproduce the observed optical/UV fluxes, although the SEDs after MJD~58120 only have one point per day (mostly in the UV filters). 

According to our modelling, the emission from the viscously heated disk does not contribute significantly to the optical and UV fluxes during the giant outburst of \swift. As a first attempt, we performed the SED fits, leaving the outer radius of the accretion disk $R_{out}$ as a free parameter. From the values providing the best fits, we observed that the fitted value of $R_{out}$  first increases until reaching value of $R_{out}\sim$0.1\,$a$ which remains constant from $\sim$MJD~58047 until the peak of the outburst. After that, the fitted value of $R_{out}$ increases from MJD~58062 to MJD~58087 reaching $R_{out}\sim$0.2\,$a$, which is roughly the Roche Lobe size of the NS. After that, the fitted value of $R_{out}$ remains constant at this value. This could be interpreted as an increase of the size of the external radius of the accretion disk and this could happens either if the accretion disk grows or dissipates. A change on the factors considering changes in the geometry of the inner flow, the fraction of absorbed emission, and the possible anisotropy of irradiation could also reproduce a similar effect. Indeed, when fixing $R_{out}\sim$0.2\,$a$ and leaving a combined factor including these effects ($k$) as a free parameter in the SED fits, the results are compatible with some of the scenarios proposed in previous works studying X-ray spectral and pulse profiles variability. 

An alternative explanation to the apparent increase of the outer radius of the accretion disk or to the change in the inner flow accretion geometry after the peak of the outburst, would be that a change in the geometry of the circumstellar disk took place. Such a change could introduce an extra contribution (more significant at redder wavelengths), but our few observational data points do not allow to disentangle between both scenarios.

The mechanisms considered in the model presented in this work are suitable to explain qualitatively the observed slopes of the optical/UV SEDs and the variability along the outburst. Under the considered assumptions, they are able to reproduce the observed fluxes. In the case of a  Be disk  not co-planar to the orbit, the irradiation of the Be disk should play a role. In that hypothetical case, not including this mechanism could lead to an overestimation of the fitted values of $R_{out}$, and would require to redefine some of the considered assumptions, such as the values of the albedo or thermalization efficiency, which could be different than the considered ones and even variable along the outburst. Correct modelling of the Be disk structure and geometry would require complex theoretical considerations and additional observations, and such a detailed analysis is proposed for a future work.

\begin{acknowledgements}
The authors thank the anonymous referee for the helpful comments and constructive remarks that have improved the quality of the manuscript. JAG and JMMH are funded by Spanish MCIN/AEI/10.13039/501100011033 grant PID2019-107061GB-C61. JvdE acknowledges a Warwick Astrophysics prize post-doctoral fellowship made possible thanks to a generous philanthropic donation and was supported by a Lee Hysan Junior Research Fellowship awarded by St Hilda's College, Oxford during part of this work. NPMK acknowledges support from the UKSA. ARE is supported by the European Space Agency (ESA) Research Fellowship. JF acknowledges the financial support from the MCIN with funding from the European Union NextGenerationEU and Generalitat Valenciana in the call Programa de Planes Complementarios de I+D+i (PRTR 2022), project (VAL-JPAS), reference ASFAE/2022/025. JAG thanks to Aastha Parikh for her help and support in the first steps of the UVOT data analysis. This research has made use of public data from the \textit{Swift} data archive.

\end{acknowledgements}

\bibliographystyle{aa}
\bibliography{SwiftJ0243pap.bib}

\end{document}